\batchmode
\documentstyle{article}
\begin{document}
\parskip 5pt plus 1pc
\parindent=16pt
\parskip 5pt plus 1pc
\parindent=16pt
\thispagestyle{empty}
\begin{flushright}
{\large AS-ITP-93-14} \\
\today
\end{flushright}
\vspace{10ex}
\centerline{\Large Modules Over Affine Lie Superalgebras}
\vspace{10ex}
\centerline{\large {\sc Jiang-Bei Fan}
and {\sc Ming YU}}
\vspace{2ex}
\centerline{\it Institute of Theoretical Physics, Academia Sinica}
\centerline{\it P.O.Box 2735, Beijing 100080, P.R.China}
\centerline{\it Fax: 086-1-2562587}
\vspace{10ex}
\centerline{\large \it Abstract}
\vspace{1ex}
\begin{center}
\begin{minipage}{130mm}
Modules over affine Lie superalgebras ${\cal G}$ are studied, in particular,
for ${\cal G}=\widehat{OSP(1,2)}$. It is shown that on studying Verma
modules, much of the results in Kac-Moody algebra can be generalized
to the super case. Of most importance are the generalized Kac-Kazhdan
formula and the Malikov-Feigin-Fuchs construction, which give the
weights and the explicit form of the singular vectors in the Verma module
over affine Kac-Moody superalgebras.
We have also considered the decomposition of the admissible representation
of $\widehat{OSP(1,2)}$ into that of $\widehat{SL(2)}\otimes$Virasoro algebra,
through which we get the modular transformations on the torus
and the fusion rules. Different boundary conditions on the torus correspond
to the different modings of the current superalgebra and characters or
super-characters, which might be relevant to the Hamiltonian reduction
resulting in Neveu-Schwarz or Ramond superconformal algebras.
Finally, the Felder BRST complex, which consists of Wakimoto modules by the
free field realization, is constructed.
\end{minipage}
\end{center}
\vfill
\hrule
\newcommand{\lsa}{Lie superalgebra}
\newcommand{\sa}{superalgebra}
\newcommand{\lsas}{Lie superalgebras}
\newcommand{\sas}{superalgebras}
\newcommand{\beq}{\begin{eqnarray}}
\newcommand{\eeq}{\end{eqnarray}}
\newcommand{\beqn}{\begin{equation}}
\newcommand{\eeqn}{\end{equation}}
\newcommand{\np}[4]{Nucl.\ Phys.\ #1 {\bf #2}, #3(#4)}
\newcommand{\cmp}[3]{Commun.\ Math.\ Phys.\ {\bf #1}, #2(#3)}
\newcommand{\pl}[4]{Phys.\ Lett.\ #1 {\bf #2}, #3(#4)}
\newcommand{\af}{\alpha}
\newcommand{\bt}{\beta}
\newcommand{\r}{\gamma}
\newcommand{\p}{\rho}
\newcommand{\k}{\kappa}
\newcommand{\0}{\theta}
\newcommand{\sg}{\sigma}
\newcommand{\h}[1]{H^ {\frac {\infty}{2} +#1}}
\newcommand{\vt}{\vartheta}
\newcommand{\df}{\partial}
\newcommand{\td}{\tilde}
\newcommand{\os}{\widehat{OSP(1, 2)}}
\newcommand{\rar}{\rightarrow}

\newtheorem{pp}{Proposition}
\newtheorem{tm}{Theorem}
\newtheorem{lm}{Lemma}
\newtheorem{corollary}{Corollary}
\newcommand{\bpp}{\begin{pp}}
\newcommand{\epp}{\end{pp}}
\newcommand{\btm}{\begin{tm}}
\newcommand{\etm}{\end{tm}}
\newcommand{\blm}{\begin{lm}}
\newcommand{\elm}{\end{lm}}
\newcommand{\bc}{\begin{corollary}}
\newcommand{\ec}{\end{corollary}}
\newcommand{\thf}[3]{\vartheta \left[ \begin{array}{c} #1\\0 \end{array}
\right]
(#2,#3)}
\newcommand{\no}{\nonumber}
\newcommand{\script}{\scriptstyle}
\newcommand{\scriptscript}{\scriptscriptstyle}
\newcommand{\crossst}{\searrow \hspace{-1em}\swarrow}
\newcommand{\crossnd}{\nearrow \hspace{-1em}\nwarrow}
\newcommand{\crossrd}{\searrow \hspace{-1em}\nearrow}
\newcommand{\crossth}{\nwarrow \hspace{-1em}\swarrow}

\newpage
\section{Introduction}
\pagenumbering{arabic}
Some recent work has been focused on the Hamiltonian reduction of the
super-group valued WZNW theory which gives rise to a super-Toda and (extended)
superconformal field theory \cite{II,Z,DGN,IMP1,IMP2}. Such procedures
might be the only accessible
way of quantizing the super-Liouville field theory, since their discretized
version, i.e.\ the super-symmetrized matrix model does not exist yet.
In ref.\cite{AGS,HY1,HY2},
it has been shown that by combining the matter sector with the
Liouville sector in a non-critical string theory one obtains a $2D$ topological
field theory, which is equivalent to the $SL(2,R)/SL(2,R)$ gauged WZNW model.
A striking feature of the non-critical string theory, as
well as the $G/G$ model, is the appearance of
infinitely many copies of the physical states with non-standard ghost numbers
\cite{LZ3,LZ4,BMP1,BMP4}.
We expect similar structure exists when $G$ is a supergroup. Namely, there must
be a close relation between the non-critical fermionic string and the $G/G$
gauged supergroup valued WZNW theory. The present paper is a preparation
toward such a consideration.

The essential ingredients of the WZNW theory is encoded in its current
algebra, the Kac-Moody algebra. It is also clear that it is the structure of
the
algebra modules that determine the physical states in $G/G$ WZNW theory.
The general structure of the affine Kac-Moody modules has been extensively
studied \cite{KK,K3,FFr,FF}.
 However, concerning its generalization, the contragradient super-algebra
module, less results can be found in the literature \cite{PR,SNR,K0,K2}.
 In this paper, we try to shed some light on the structure of modules over
superalgebras,
which might be relevant to our understanding of the (gauged) supergroup
valued WZNW theory.
As a consequent problem, the analysis of
the BRST semi-infinite cohomology of the $G/G$ WZNW theory,
where $G$ is in general a supergroup, will be done
in a separate paper \cite{FY}.
Our main results are the generalized
Kac-Kazhdan formula, eq.(\ref{5.100}) and the generalized MFF
\cite{MFF} construction.
 To give a clear idea what we have done in this paper, let us recall some
essential facts about our knowledge of the infinite dimensional Lie algebras.
In \cite{KK}, the structure of Verma module over an infinite dimensional Lie
 algebra was studied; Kac-Kazhdan formula tells whether a Verma module is
reducible, and gives the weights of the singular vectors in a Verma module.
In \cite{MFF}, the explicit form of such a vector in a Verma module
was constructed, in a way we might call it MFF
construction. Moreover, the Wakimoto modules (namely,
when restricted to Virasoro algebra, the Feigin-Fuchs modules \cite{FFu}),
were extensively studied \cite{FFr,FF}.

Our paper is organized as follows.
In section \ref{sec-\lsa}, we review some fundamental knowledge  about
\lsa. In section \ref{sec-Character}, we study a hidden Virasoro algebra in
${\cal U}(\widehat{OSP(1, 2)})$ through
GKO construction \cite{GKO1,GKO2} of
$\widehat{OSP(1, 2)} /\widehat{SL(2)} $.
We get the decomposition of the representation of
$\widehat{OSP(1, 2)}$ into $\widehat{SL(2)}\otimes$
Virasoro, and concentrate on
the so called admissible representations \cite{KW,FKW}. Later we deal
with the $S$ modular transformation and the fusion rules. Fusion rules of
the the admissible representations are given through Verlinde formula \cite
{V,MS,AG}. Similar to those of $\widehat{SL(2)}$, there are negatives
integers appearing in the fusion rules,
which might be explained as that there are lowest weight states
appearing in the fusion of two highest weight states. It should be noted
that the decomposition, eq. (\ref{4.200}), is first given in \cite{KW}
in a sophisticated
way. We get the $S$-modular transformation of
$\widehat{OSP(1, 2)}$ through that of
the Virasoro and $\widehat{SL(2)}$, which are already known.
The characters of the admissible $\widehat{OSP(1, 2)}$-modules
are obtained by combining
that of Virasoro and $\widehat{SL(2)}$.

In section \ref{sec-Structure}, we study the $\widehat{OSP(1, 2)}$-module
in general, and give out the
structure of all $\widehat{OSP(1, 2)}$-modules. The result is analogous to the
classification of Feigin and Fuchs on Virasoro-module \cite{FFu} and that of
Feigin-Frenkel on $\widehat{SL(2)}$-module \cite{FF}.
The Kac-Kazhdan
formula \cite{KK} and MFF construction \cite{MFF} is generalized
to superalgebra, as it had been mentioned above.

Section 5 is devoted to the construction of Wakimoto modules \cite{Wo},
which is the free field realization of the current superalgebra \cite{BO1}.
The Felder cohomology \cite{Fl,BF} is analyzed in detail for ${\cal U(G)}=\os$.

In conclusion, we speculate that for a general affine superalgebra $\cal G$,
the coset space construction for ${\cal G/G}_0$, where ${\cal G}_0$ is the
even part of $\cal G$, will result in a $W_N$ algebra. In appendix we deal
with some technical details.

\section{\lsa \  and Current \sa}
\label{sec-\lsa}
In this section we review some fundamental properties
of the \lsas, mainly to make our paper self-contained.
Detailed discussion can be found in \cite{PR,K0,K2}.
The representation theory of $OSP(1, 2)$ has been studied by the
authors of ref.\cite{PR,SNR}.

\subsection{\lsa}
\lsa, namely, graded Lie algebra in mathematical terms, can be written as
${\cal G=G}_{\bar{0}}+{\cal G}_{\bar{1}}$. ${\cal G}_{\bar{0}}$,
the even part of $\cal G$, is by itself a Lie algebra.

Denote the generators of $\cal G$ by $\tau^{\alpha}$, $\alpha =1,\ldots,
d_{{\cal G}_{\bar{0}}}+d_{{\cal G}_{\bar{1}}}
(\mbox {or }\infty, \mbox{when }\ {\cal G} \ \mbox {infinite dim. })$,
 where $d_{{\cal G}_{\bar{i}}}$
is the dimension of ${\cal G}_{\bar{i}}$. The commutators are
\beq
[\tau^{\af}, \tau^{\bt}]=f^{\af \bt}_{\ \ \r}\tau^{\r},
\eeq
where $f^{\af \bt}_{\ \ \r}$ are the structure constants. The Lie
bracket $[\ ,\ ]$ is defined to be
\beqn
[a,b]=ab-(-1)^{deg(a)deg(b)}ba
,\eeqn
where $deg(a)=0$(resp. 1) for $a\in {\cal G}_{\bar{0}}
(\mbox{resp. }{\cal G}_{\bar{1}})$.
 From Jacobi identity
\beq
[\tau^{\af}, [\tau^{\bt}, \tau^{\r}]=[[\tau^{\af}, \tau^{\bt}], \tau^{\r}]
+(-1)^{d(\af)d(\bt)}[\tau^{\bt}, [\tau^{\af}, \tau^{\r}]],
\eeq
we get
\beq
f^{\af \p}_{\ \ \sg}f^{\bt \r}_{\ \ \p}=f^{\af \bt}_{\ \ \p}
f^{\p \r}_{\ \ \sg}+
(-1)^{d(\af)d(\bt)}f^{\bt \p}_{\ \ \sg}f^{\af \r}_{\ \ \p}
\eeq
where $d(\af)=0$(resp. 1) for $\tau^{\af}\in {\cal G}_{\bar{0}}
(\mbox{resp. }{\cal G}_{\bar{1}})$.

In \cite{K2}, Kac gave out a classification of all finite growth contragradient
\lsas, each associated with a generalized indecomposable $n\times n$
Cartan matrix and a subset of $\{1\ldots n \}$.

Let $A=(a_{ij})$ be an $n\times n$ generalized Cartan matrix;
$\tau \subseteq I=\{1 \ldots n\}$. $a_{ij}$ satisfy
\beqn\begin{array}{ll}
  a_{ii}=2,\ \ \ a_{ij}\leq 0
  &\forall i\neq j,\ \ \  i, j\in I,  \\
  a_{ij}=0\leftrightarrow a_{ji}=0. &
  a_{ij} \mbox{ is even} \ \ \ \forall i\in \tau.\end{array}
\eeqn
The Chevalley basis $\{e_{i}, f_{i}, h_{i}, i=1\ldots n \}$ satisfy
\beqn \begin{array}{ll}
  [e_{i}, f_{j}]=\delta _{ij} h_{i};&
  {[}h_{i}, h_{j}]=0;\\
{[}h_{i}, e_{j}]=a_{ij} e_{j};&
{[}h_{i}, f_{j}]=-a_{ij} f_{j};
\end{array}
\eeqn
\beqn
(ade_{i})^{-a_{ij}+1} e_{j}=(adf_{i})^{-a_{ij}+1}f_{j}=0\ \ \ \forall i\neq j,
\ \ \ \ i, j\in I;
\eeqn
where the $Z_{2}$-degrees of $e_{i}, f_{i}, h_{i}$ are
\beqn\begin{array} {ll}
deg(h_{i})=0, &\forall i\in I;\\
deg(e_{i})=deg(f_{i})=0, & \forall i\not\in \tau;\\
deg(e_{i})=deg(f_{i})=1, &\forall i\in \tau. \end{array}
\eeqn

{\it Remark}: (super)Virasoro algebra is not a contragradient superalgebra.

\subsection{Finite dim. \lsas}
In this subsection, we shall restrict our discussion the cases that
dim$(\cal G)$ is finite.
Similar to Lie algebra, the adjoint representation of the
finite dimensional Lie superalgebra takes the form
\beq
(F^{\af})^{\bt}_{\ \r}=f^{\af\bt}_{\ \ \ \ \r}. \label{9331901}
\eeq
Define the supertrace,
\beq
str(F)=\sum_{\af} (-1)^{d(\af)}F^{\af}_{\ \ \af}.
\eeq
There exists a metric tensor
\beq
h^{\af\bt}=str(F^{\af}F^{\bt}), \label{9331902}
\eeq
with relation
\beq
h^{\af\bt}=(-1)^{d(\af)d(\bt)}h^{\bt\af. }
\eeq
We would like to restrict our discussion to the semisimple Lie
superalgebra, (which follows from the definition in ref. \cite{PR}), i. e.
${\cal G}_{\bar{0}}$ is semisimple and
\beq
det|h^{\af\bt}|\neq0.
\eeq
Define the inverse of h
\beq
h_{\af\bt}h^{\bt\r}=\delta_{\af}^{\ \r}.
\eeq
It can be verified \cite{PR} that
\beq
C=\tau^{\af}h_{\af\bt}\tau^{\bt}
\label{2.10}
\eeq
is a Casimir operator. In the adjoint representation
\beq
C_{ad}={\bf 1}.
\eeq

To make our discussion more concrete we consider a simple example,
the $OSP(1, 2)$ superalgebra,
\beqn\begin{array}{ll}
\{ j^{+}, j^{-} \}=2J^{3};& \{j^{\pm}, j^{\pm}\}=\pm2J^{\pm};  \\
{[}J^{3}, j^{\pm}]=\pm \frac{1}{2}j^{\pm};&[J^{\pm}, j^{\mp}]=-j^{\pm};  \\
{[}J^{+}, J^{-}]=2J^{3};&[J^{3}, J^{\pm}]=\pm J^{\pm}, \end{array}
\eeqn
while other (anti)commutators vanish. We see that the even generators,
$J^{\pm}, J^{3}$ constitute a $SL(2)$ subalgebra.
The irreducible representation
of $OSP(1, 2)$ with highest weight $j$ (except for $j=0$)
can be decomposed  into the irreducible representations of $SL(2)$,
with isospin $j$ and $j-\frac{1}{2}$ respectively.
The HWS of the $OSP(1, 2)$ is defined as
\beqn
j^{+}|j, j, j\rangle =0;\ \ \ \
J^{3}|j, j, j\rangle =j|j, j, j\rangle , \label{9331701}
\eeqn
where the first index in $|i,j,k\rangle$ labels the $\os$ representation
and the last two indices refer to the $SL(2)$ isospin and its third component.
By eq.(\ref{9331701}), $|j,j,j\rangle$ is also a HWS for $SL(2)$ with isospin
$j$.
Then
\beqn
|j, j-1/2, j-1/2\rangle =j^{-}|j, j, j\rangle
\eeqn
satisfies
\beqn
J^{+}|j, j-1/2, j-1/2\rangle =0,
\eeqn
and generates another $SL(2)$ irreducible module of highest weight $j-1/2$.
The lowest dimensional faithful representation of $OSP(1,2)$
is given by $j=1/2$, for which
\beqn
\begin{array}{ll}
str\{J^+J^-\}=str\{J^-J^+\}=1;&str\{j^+j^-\}=-str\{j^-j^+\}=2;\\
str\{J^3J^3\}=1/2.& \end{array}
\eeqn
The adjoint representation is of $j=1$, by eqs.(\ref{9331901},\ref{9331902})
the metric can be worked out explicitly,
\beqn\begin{array}{ll}
h^{+-}=h^{-+}=3;&h^{\frac{1}{2},-\frac{1}{2}}=-h^{\frac{1}{2},-\frac{1}{2}}=6;
\\h^{33}=3/2.&\end{array}
\eeqn
By eq.(\ref{2.10}), the second Casimir operator of $OSP(1,2)$ is
\beqn
C=1/3\{J^{+}J^{-}+J^{-}J^{+}+2J^{3}J^{3}-\frac{1}{2}j^{+}j^{-}+
\frac{1}{2}j^{-}j^{+}\}.
\eeqn
For an irreducible representation labeled by  isospin $j$
\beqn
C=j(2j+1)/3\cdot{\bf 1}.
\eeqn
\subsection {Current \sas}
We now turn to the current superalgebra $J^{\af}$ associated with a
finite dimensional \lsa\ {\cal G}. The currents satisfy the short
distance operator
product expansion ( OPE ),
\beq
J^{\af}(z_1)J^{\bt}(z_{2})=\frac{\td{k}h^{\af\bt}}{z_{12}^{2}}+
\frac{f^{\af\bt}_{\ \ \r}J^{\r}(z_{2})}{z_{12}}.
\label{2.20}
\eeq
By Sugawara construction,
 we get the stress-energy tensor
\beq
T(z)=\frac {:J^{\af}h_{\af \bt}J^{\bt}:}{2\td{k}+1}, ~~~~~
c=\frac {2\td{k}\, sdim({\cal G})}{2\td{k}+1},
\eeq
 where $c$ is the central extension of the Virasoro algebra and
\beq
sdim({\cal G})=d_{{\cal G}_{\bar{0}}}-d_{{\cal G}_{\bar{1}}}.
\eeq
The level of the current superalgebra, eq.(\ref{2.20}), is
\beqn
k=2\td{k}\td{h}_G,
\eeqn
where $\td{h}_G$ is the dual Coexter number for Lie algebra, $3/2$
for $OSP(1,2)$. For an integrable representation of $\os$, $k$ is an integer.

\section{(Super)characters and Modular Transformations}
\label{sec-Character}
The affine Kac-Moody superalgebra associated with $OSP(1,2)$, eq.(\ref{2.20}),
assumes the following form,
\beq
&\begin{array}{ll}
 \{ j^{+}_r, j^{-}_s \}=2J^{3}_{r+s}+2rk\delta_{r+s,0};
 & \{j^{\pm}_r, j^{\pm}_s\}=\pm2J^{\pm}_{r+s};  \\
 {[}J^{3}_n, j^{\pm}_r]=\pm \frac{1}{2}j^{\pm}_{n+r};
 &[J^{\pm}_n, j^{\mp}_r]=-j^{\pm}_{n+r};  \\
 {[}J^{+}_n, J^{-}_m]=2J^{3}_{n+m}+nk\delta_{n+m,0};
 &[J^{3}_n, J^{\pm}_m]=\pm J^{\pm}_{n+m}, \end{array}\\
\label{s4001}
                 &{[}d,J_n^\af]=nJ_n^\af. \no
\eeq
Here, in general, the odd generators could be integral or half-integral
moded. However, in the case of $\os$, it is clear that the two modings
are isomorphic. Unless specified we shall assume the integral moding for
all the generators. In subsection (3.3) we shall come to this point again.

The highest weight state of $\os$ is an eigenstate of $J_0^3$ and
annihilated by $J_n^a,\ j_n^\af,\ n\in N$ and $j_0^+$.
The highest weight representation of $\os$ is generated by acting on
the highest weight state with the lowering operators. It is useful to
notice that the even part of $\os$ is itself a $\widehat{SL(2)}$ Kac-Moody
algebra. So in general we can decompose the representation space of $\os$
into a tensor product of two spaces,
$$\os \sim \widehat{SL(2)} \otimes \os/ \widehat{SL(2)}.$$
Such decomposition is carried out explicitly in subsection (3.1). Similarly,
the (super) characters of $\os$ can be written in terms of that of
$\widehat{SL(2)}$ and its branching functions, as what is done in subsection
(3.2). In section 4 we shall show that these results can be rederived
from the pure algebraic approach.

\subsection{The Decomposition}
The even part of $\os$ is the $\widehat{SL(2)}$ subalgebra.
The energy momentum tensor for $SL(2)$ WZNW theory is, via Sugawara
construction,
 \beq
T^{SL(2)}(z)=\frac{:J^{3}J^{3}+\frac{1}{2}J^{+}J^{-}
+\frac{1}{2}J^{-}J^{+}:}{k+2}.
\eeq
By GKO\cite{GKO1,GKO2} construction,
we get a Virasoro algebra for the coset space $V = OSP(1,2)/SL(2)$,
\beq
T^{V}(z)=T^{OSP(1, 2)}(z)-T^{SL(2)}(z).
\eeq
It is easy to verify that
\beqn  \begin{array}{l}
[T^{V}, T^{SL(2)}]=0;\\
c^{OSP(1,2)}= \frac{2k}{2k+3};\\
c^{SL(2)}= \frac{3k}{k+2};\\
c^{V}=1-6\frac{(k+1)^{2}}{(2k+3)(k+2)}. \end{array}
\label{4.10}
\eeqn
So far we are interested in the rational conformal field theory(RCFT), for
which
the possible sets of the characters can be classified.
For $\widehat {SL(2)}_{k}$ RCFT \cite{KW,MP},
the level $k$ satisfies
\beq
k+2=\frac{p}{q}, \label{4.20}
\eeq
where $p, q$ are coprime positive integers. Then by eqs.(\ref{4.10},
\ref{4.20})
\beq
c^{V}=1-6\frac{(p-\td{q})^{2}}{p \td{q}},
\eeq
where
\beq
\td{q}=2p-q.
\eeq
We see that $gcd(p, \td{q})=1$. So the Virasoro algebra for the coset space
$V=OSP(1,2)/SL(2)$ is
in the $(p, \td{q})$ minimal series, provided
\beq
p, q, 2p-q>0.
\eeq
Similar consideration has been taken by Kac and Wakimoto \cite{KW}
in the study of the admissible representation
of $\widehat{OSP(1, 2)}$.
Notice that the unitarity of $\widehat{SL(2)}$  demands $q=1,\ p\geq 2$,
which leads to a non-unitary Virasoro minimal series in the $V$ sector
except for the trivial case, $c^V=0$.

{\it Remark}: $\os$ WZNW theory is not a unitary theory except for the trivial
case, $k=0$.

The admissible representation of $\widehat{SL(2)}$ \cite{KW,MP} is
classified as
\beq
2j+1=r-s \frac{p}{q};
r=1, \ldots , p-1&s=0, \ldots, q-1, \label{4.30}
 \eeq
where $j$ is the isospin of the HWS.
Now we consider an HWS in $\widehat{OSP(1, 2)}$ with the level
and the isospin as in eqs.(\ref{4.20}, \ref{4.30}).
It is easy to see that such a HWS is a tensor product of the HWS's in
$\widehat{SL(2)}$ and $V$ sector separately,
$$|HWS\rangle_{\os}=|HWS\rangle_{\widehat{SL(2)}}\otimes
|HWS\rangle_{V}$$
Its conformal weight in the $V$ sector
is
\beq
h_{V}&=&h_{\os}-h_{\widehat{SL(2)}}\no \\
&=&\frac{j(j+1/2)}{k+3/2}-\frac{j(j+1)}{k+2}=\frac{j(j-k-1)}{(2k+3)(k+2)}
   \no \\
&=&\frac{(mp-r\td{q})^{2}-(p-\td{q})^{2}}{4p\td{q}}  \no\\
&=&h_{r, m},
\eeq
where
\beqn
m=2r-s-1, \label{4.40}
\eeqn
Again we find that $h_{V}$ is indeed in the set of
conformal weights of the
minimal $(p, \td{q})$ Virasoro series provided we restrict
the possible values of $r, s$ to be
\beqn
0<r<p,\ \ \ 0<m<\td{q}.
\eeqn
Soon after we shall see that upon this restriction
the super-characters form a representation of the modular group on torus.

To decompose an irreducible $\os$ module,
let us proceed to find out all the states $|\phi \rangle$
in the $\widehat{OSP(1, 2)}$ module
which are HWS's for both $\widehat{SL(2)}$
and Virasoro $(p,\td{q})$ algebra,
\beqn
|\phi\rangle=|\td{r},\td{s}\rangle_{SL(2)}\otimes|l,\td{m}\rangle_{V}.
\eeqn
Assuming that such a state appears in the Nth level of the $\os$ module, we try
to solve the following equations,
\beqn \left\{ \begin{array}{lll}
\frac{j(j+1/2)}{k+3/2}+N&=&\frac{\td{j}(\td{j}+1)}{k+2}+h_{l, \td{m}},
N\in Z_{+}, \\
\td{j}=\frac{1}{2}(\td{r}-1)-\frac{1}{2}\td{s}\frac{p}{q}
&=&j+n/2, n\in Z, \end{array}\right.
\eeqn
which has infinitely many solutions
   \beq
&&N=\frac{1}{2}n(n+1)+t^{2}p\td{q}+t(r+n)\td{q}-tmp, \\
&&\td{j}=j+n/2,~~ l=r+n+2tp,~~ \td{m}=m,\no
\eeq
or
\beq
&&N=\frac{1}{2}n(n+1)+t^{2}p\td{q}-t(r+n)\td{q}-tmp+ml, \\
&&\td{j}=j+n/2,~~ l=2tp-(r+n),~~ \td{m}=m,\no
\eeq
where $n,~t\in Z$ such that $N>0, \mbox{ or }N=0,\ n<0$.
Because we are considering the irreducible representations of $\os$,
some solution
should be excluded. We see that the state $|\pm l+2tp, s\rangle_{SL(2)}
\otimes |\pm l+2t'p, m\rangle_{V}$ is a singular state in the Verma module
$M(|l, s\rangle)_{SL(2)}\otimes M(|l, m\rangle)_{V}$, if $0\leq l<p$.
So in the decomposition we should restrict ourselves to the
irreducible $SL(2)$ and Virasoro modules.
After taking this into account, we get
\beq
L_{m, s}^{OSP(1, 2)}=\sum_{l=1}^{p-1}L_{l, s}^{SL(2)}\otimes L_{l, m}^{V}&
,~~~m+s\in \mbox{ odd}, \label{4.200}
\eeq
where $L_{m, s}^{OSP(1, 2)}$ is the irreducible reprepresentation
for $\os$ with the HWS of level k and isospin $j_{m,s}$
\beq
&& k+3/2=\frac{\td{q}}{2q},~~~~q+\td{q}\in \mbox{even},~~gcd(q,\
\frac{q+\td{q}}{2})=1;\no\\
&&4j_{m, s}+1=m-s\ \frac {\td{q}}{q},
\ \ \ \ m=1,\cdots,\td{q}-1 \ \ s=0,\cdots,q-1.
\label{4.60}
\eeq
It is easy to see that the isospin $j$ in eq.(\ref{4.60}) is exactly
that in eq.(\ref{4.30}).
However it may be convenient to adopt the ($q$ ,  $\tilde{q}$) and ($m$, $s$)
notation for the case of $\os$ (instead of the labels ($p$, $q$) and
($r$, $s$) for $\widehat{SL(2)}$).

\subsection{Characters and Supercharacters on Torus}
We note that for a $Z_{2}$ graded algebra, the module should also be
$Z_{2}$ graded in consistence with that of the algebra.
For convenience we
assume that the $Z_2$ degree of the HWS of a HW module is even.
The character $\chi$ and the supercharacter
$S\chi$ of a ${\cal G}$-module $R$ are defined as
\beq
\chi =tr_{R}e^{i2\pi\tau(L_{0}-\frac{c}{24})+i2\pi zJ_{0}^{3}};\
S\chi =str_{R}e^{i2\pi\tau(L_{0}-\frac{c}{24})+i2\pi zJ_{0}^{3}}
\eeq
The characters and supercharacters so defined correspond to the different
boundary conditions along $\tau$ direction.
Notice that up to a phase factor
the supercharacter can be obtained from the character
by letting $z \rightarrow z+1$.

Using the decomposition, eq.(\ref{4.200}), we have
\beq
&\chi_{m, s}^{OSP(1, 2)}(z,\tau)=
\sum_{l=1}^{p-1}\chi_{l, s}^{SL(2)}(z,\tau)
\chi_{l, m}^{V}(\tau), \no\\
&S\chi_{m, s}^{OSP(1, 2)}(z,\tau)=
\sum_{l=1}^{p-1}(-1)^{l+r}\chi_{l, s}^{SL(2)}(z,\tau)
\chi_{l, m}^{V}(\tau), \label{4.70} \\
&m+s \mbox{ is odd}, \  m=1,\ \ldots \td{q}-1, \ s=0,\ \ldots p-1.   \no
\eeq
$\chi_{l, m}^{V}(\tau)$'s so defined are called the
($\os\supset \widehat{SL(2)}$) branching functions.

Modular transformations of the characters of the
$\widehat{SL(2)}$ admissible representations
and Virasoro minimal models
have been
studied in detail in ref.\cite{C,IZ,CIZ,MP}. Then the
modular transformations of the $\os$ (super)characters can be obtained
by using the decomposition formula eq.(\ref{4.200}).

Under $S:\tau\rightarrow -1/\tau, z\rightarrow z/\tau$
\beq
\td {\chi}^{SL(2)}_{l, s}&=&\sum_{l'=1}^{p-1} \sum_{s'=0}^{q-1}
S^{l's', SL(2)}_{l\ s}\chi^{SL(2)}_{l's'}; \no \\
\td {\chi}^{V}_{l, m}&=&\sum_{l'=1}^{p-1} \sum_{m'=0}^{\td{q}-1}
S^{l'm', V}_{l\ m}\chi^{V}_{l'm'}; \label{4.110}
\eeq
where
\beq
S^{l's', SL(2)}_{l\ s}&=&\sqrt{\frac{2}{pq}}(-1)^{ls'+sl'}e^{-i\pi ss'p/q}
\sin{\frac{\pi qll'}{p}};\no\\
S^{l'm', V}_{l\ m}&=&\sqrt{\frac{2}{p\td{q}}}(-1)^{lm'+ml'+1}
\sin{\frac{\pi mm'p}{\td{q}}}\sin{\frac{\pi qll'}{p}}. \label{4.120}
\eeq

{}From eqs.(\ref{4.110}-\ref{4.120}), we get the $S$ modular
transformation of supercharacters
$S\chi^{\os}_{m, s}$, when $1\leq m=2r-s-1<\td{q}$, after a lengthy
calculation.
\beq
S\td {\chi}^{\os}_{m, s}&=&\sum_{m'=1}^{\td{q}-1}
\sum_{\tiny\begin{array}{c}\vspace{-1ex}s'=0,\\ \vspace{-1ex}
s'+m'\in odd\end{array}}^{q-1}
S^{m's', \os}_{m\ s}S\chi^{\os}_{m's'};\no\\
S^{m's', \os}_{m\ s}&=&\sqrt{\frac{2}{p\td{q}}}(-1)^{(m-s+s'-m')/2}e^{-i\pi
ss'p/q}
\sin{(\pi qmm'/\td{q})}.\label{9331910}
\eeq
We see that when $1\leq m<2p-q=\td {q}, 0\leq s<q,
m+s \in odd,\ S\chi^{\os}_{m, s}$
form a representation of the modular group. So we get a rational
conformal field theory. On the contrary,  the characters are not
closed under $S$ modular transformations. This issue
will be
addressed again in the next subsection.

Generally one can figure out the fusion rules through the $S$ matrix
\cite{V,AG},
The relation between the fusion rules and the $S$ matrix is conjectured
by Verlinde \cite{V}, later proved in \cite{MS}.
\beq
N_{ij}^{\ \ k}=\sum_n \frac{S_{i}^{\ n}S_{j}^{\ n}S_{k}^{*n}}{S_0^{\ n}}
\label{4.140}
\eeq
However from the relation eq.(\ref{4.140}), we get $-1$ for some $N_{ij}^{
k}$'s,
as it also happens
in the case of $\widehat{SL(2)}$ \cite{BF,AY}. This can be interpreted as
that there are
lowest weight states appearing in the fusion of two HWS's. The fusion rules
so obtained are

\begin{enumerate}
 \item when $s_{1}+s_{2}<q$
   \beq
     [\varphi^{\os}_{m_{1}, s_{1}}][\varphi^{\os}_{m_{2}, s_{2}}]=
\sum_{\tiny\begin{array}{c}\vspace{-1ex}m_{3}=|m_{1}-m_{2}|+1\\
\vspace{-1ex}m_3+m_1+m_2
\in odd\end{array}}
^{min[m_{1}+m_{2}, 2\td{q}-m_{1}-m_{2}]-1}
[\varphi^{\os}_{m_{3}, s_{1}+s_{2}}];
\eeq
 \item when $s_{1}+s_{2} \geq q$
   \beq
     [\varphi^{\os}_{m_{1}, s_{1}}][\varphi^{\os}_{m_{2}, s_{2}}]=
\sum_{\tiny\begin{array}{c}\vspace{-1ex}m_{3}=|m_{1}-m_{2}|+1,\\
\vspace{-1ex}m_3+m_1+m_2 \in odd
\end{array}}
^{min[m_{1}+m_{2}, 2\td{q}-m_{1}-m_{2}]-1}
[\td {\varphi}^{\os}_{m_{3}, s_{1}+s_{2}-q}].
\eeq
 \end {enumerate}
Here, $[\td{\varphi}^{\os}_{m, s}]$ denotes the conformal block
corresponding to the lowest weight representation.

The $T$ matrix for an
irreducible module under : $\tau\rar\tau+1$,
is diagonal,
\beqn
S\chi \rar e^{i\pi/4(\frac{(4j+1)^2}{2k+3}-1)}S\chi.
\eeqn
Here $j$ and $k$ are the isospin and the level resp.. For the admissible
representations(c.f. eq.(\ref{4.60})), we rewrite the matrix element as,
\beq
&T^{m',s'}_{m\ ,s}=e^{i\pi/4(\frac{(qm-\td{q}s)^2}{q\td{q}}-1)}
\delta_{m,m'}\delta{s,s'}\\
&m'+s',\ m+s \in \mbox{odd}.\no
\eeq
It is easy to verify that the following identities hold,
\beqn
S^*S=T^*T=S^4=T^{4[q,\td{q}]}=(ST)^3={\bf 1},
\eeqn
where $[q,\td{q}]$, defined as the least common multiple of $q,\ \td{q}$,
equals $q\td{q}\ (q\td{q}/2$)\ ,when $gcd(q,\td{q})
=1\ ( 2$).

In \cite {KW,MP}, the authors pointed out that the characters of $\widehat
{SL(2)}$ ($s\neq 0$) have
a simple pole at $z=0$, and that the residue contains a Virasoro characters
as a factor. However, here we find that the supercharacters of
$\os$ for the admissible representations are regular while
the characters singular at $z=0$. In fact, this is due to different locations
of the poles of the modular functions in the $z$ complex plane.

Now let us calculate the characters of the  $\os$ modules in a more explicit
form. It is useful to consider the following $\vartheta$ function,
\beqn
\thf{\frac{b}{a}}{z}{a\tau}=\sum_{n\in Z}e^{i\pi a\tau(n+b/a)^2+i2\pi z
(n+b/a)}
\eeqn
The following product expansion of two $\theta$ functions \cite{M,KY}
is used many times in our calculation,
\beq
&\thf{\frac{a_1}{n_1}}{z_1}{n_1 \tau}
\thf{\frac{a_2}{n_2}}{z_2}{n_2 \tau}=\sum_{d \in Z_{n_1+n_2}}\no \\
&\thf{\frac{n_1d+n_1+n_2}{n_1+n_2}}{z_1+z_2}{(n_1+n_2)\tau}
\thf{\frac{n_1n_2d+n_2a_1-n_1a_2}{n_1n_2(n_1+n_2)}}
{n_2z_1-n_1z_2}{n_1n_2(n_1+n_2)\tau}
,
\label{100}
\eeq
from which we get
\beq
\left\{\thf{1/6}{3z/2}{3\tau}-\thf{1/6}{3z/2}{3\tau}\right\}
\thf{1/2}{z/2}{\tau}=\no\\
e^{i\pi/3}\thf{1/2}{z+1/2}{\tau}\thf{1/6}{1/2}{3\tau}.\label{101}
\eeq
For a  $\os$ Verma module labeled by isospin j and level k, we have
\beq
\chi M_{j}&=&e^{i2\pi z(j+1/4)+i2\pi \tau \frac{(j+1/4)^{2}}{k+3/2}}
\left\{ \thf{1/6}{3/2z}{3 \tau}- \thf{-1/6}{3/2z}{3 \tau}\right\}^{-1},\no\\
S\chi M_{j}&=&e^{i \pi/2+i2\pi z(j+1/4)+i2\pi \tau \frac{(j+1/4)^{2}}{k+3/2}}
\\
&&\left\{ \thf{1/6}{3/2z+3/2}{3 \tau}- \thf{-1/6}{3/2z+3/2}{3 \tau}
\right\}^{-1}.\no
\eeq
\bpp
For the admissible representations $L_{m,s}^{\os}$,
\beq
\chi_{m,s}^{\os}&=&e^{-i/3\pi} \thf{1/2}{z/2}{\tau}
\left\{ \thf{1/6}{1/2}{3\tau}\thf{1/2}{z-1/2}{\tau}\right\}^{-1}
\no\\
   & &
\left\{ \thf{\frac {qm-\td{q}s}{2q\td{q}}}{1/2\td{q} z}{q \td{q}\tau}-
  \thf{\frac {-qm-\td{q}s}{2q\td{q}}}{1/2\td{q} z}{q \td{q}\tau} \right\},\no\\
\label{9331702}
S\chi_{m,s}^{\os}&=&e^{-i/3\pi} \thf{1/2}{\frac{z+1}{2}}{\tau}
 \left\{ \thf{1/6}{1/2}{3\tau}\thf{1/2}{z+1/2}{\tau}\right\}^{-1}
   \no\\
   & &
\left\{ \thf{\frac {qm-\td{q}s}{2q\td{q}}}{\frac{1}{2}\td{q} (z+1)}
{q \td{q}\tau}-
  \thf{\frac {-qm-\td{q}s}{2q\td{q}}}{\frac{1}{2}\td{q} (z+1)}{q \td{q}\tau}
\right\}.
 \eeq
\epp
Proof. We need only to prove the equation valid for the character.
Using eq.(\ref{100}) and the character formula for $\widehat{SL(2)}$ and
Virasoro algebra \cite{CIZ}
\beq
\chi^{V}_{l,m}=\frac{\thf{\frac{\td{q} l-pm}{2p\td{q}}}{0}{2p\td{q}\tau}-
\thf{\frac{\td{q} l+pm}{2p\td{q}}}{0}{2p\td{q}\tau}}{\eta(\tau)}
 \label{4.80}\\
\chi^{SL(2)}_{l,s}=\frac{\thf{\frac{ql-ps}{2pq}}{pz}{2pq\tau}-
\thf{\frac{ql+ps}{2pq}}{pz}{2pq\tau}}{\pi (\tau,z)} \no
\eeq
where
\beq
\eta(\tau)=q^{1/24}\prod(1-q^n)=e^{-i\pi/6}\thf{1/6}{1/2}{3\tau},
\ \mbox{for}\ q=e^{i2\pi\tau};\no\\
\pi(\tau,z)=e^{-i\pi/2}\thf{1/2}{z+1/2}{\tau}=
e^{i\pi/2}\thf{1/2}{z-1/2}{\tau}.
\eeq
Moreover, when $l=q,0$, or $m=\td{q},0$,
\beq
\thf{\frac{\td{q} l-pm}{2p\td{q}}}{0}{2p\td{q}\tau}-
\thf{\frac{\td{q} l+pm}{2p\td{q}}}{0}{2p\td{q}\tau}=0.
\eeq
Using eqs.(\ref{100},\ref{101}), the character formula, eq.(\ref{9331702}),
is obtained.

We see that $z=0$ is a first order zero of the functions
$\thf{1/2}{(z+1)/2}{\tau}$ and  $\thf{1/2}{z+1/2}{\tau}$, so
the supercharacter of $\os$ is not singular as $z \rightarrow 0$.

{}From the character of the Verma module and that of the irreducible module
we can draw some information about the structure of these module.
Notice that
\beq
\frac{\chi L^{\os}_{m,s}}{\chi M^{\os}_{m,s}}=
\sum_n e^{i\pi\tau(q\td{q}n^2+n(qm-\td{q}s))+i\pi\td{q}zn}-
\sum_n e^{i\pi\tau(q\td{q}n^2-n(qm+\td{q}s+ms))+i\pi z(\td{q}n-m)}
.\eeq
The equation is very similar to that of $\widehat{SL(2)}$ and Virasoro
minimal series
\cite{R,KW,MP,BF}.
We may conjecture that the singular vectors in the admissible representations
are of isospin $4j+1=2\td{q}n\pm m-s\frac{\td{q}}{q}$.
Indeed this can be verified in the section 4.

\subsection{NS and R type (Super-)Characters}
Similar to the super-Virasoro algebras, there are two types of $\os$ algebras.
One is of Neveu-Schwarz( NS ) type, another of Ramond( R ) type.
The two types of $\os$ are in fact isomorphic while the corresponding Virasoro
algebras by Sugawara constructions are somewhat different.
The generators of R type
$\os$ are all integer moded while the fermionic generators of NS type
$\os$ are half integer moded. Physically, the difference corresponds
to the different boundary conditions, which might be seen from the modular
transformations of the characters and the supercharacters. The subscript $r$ of
$j_r^{\pm}$ in eq.(\ref{s4001}) satisfies
$r\in {\cal Z}({\cal Z}+1/2,$ resp.) for R type
(NS type, resp.) $\os$. There exists an isomorphism map from R type $\os$
to NS type $\os$
\beqn
\begin{array}{cc}
j^{\pm 1/2}_n\rar \pm j_{n\pm1/2}^{\mp1/2};&J_n^{\pm}\rar J_{n\pm1}^{\mp};\\
J_n^3\rar-J_n^3+k/2\delta_{n,0};&k\rar k;\\
d\rar d+J_0^3,&
\end{array}
\label{9331703}\eeqn
which sends ${\cal G}_{\pm}$ to ${\cal G}_{\pm}$, ${\cal G}_0$ to
${\cal G}_0$.

For convenience, we add a superscript NS or R on the generators to
 distinguish the two types of superalgebras, for example,
\beqn
J_0^{3,NS}=\frac{k}{2}-J_0^{3,R}.
\label{001}
\eeqn
The Virasoro generators by Sugawara construction, $\{ L_n^{NS}\}$
and $\{ L_n^{R}\}$, are also related,
\beq
L_0^{R}&=&L_0^{NS}-J_0^{3,NS}+k/4;\no\\
L_n^{R}&=&L_n^{NS}-J_n^{3,NS},\ n\neq 0.
\label{002}
\eeq
By the isomorphism, eq.(\ref{9331703}), a highest weight module over
R type $\os$ is mapped to a highest weight module
over NS type $\os$. In the following discussion, we adopt the notation
in the previous subsection except for the superscript NS or R.
Now we have two kinds of characters $\chi ^{NS}_{V},\chi ^{R}_{V}$
and supercharacters $S\chi ^{NS}_{V}$, $S\chi ^{R}_{V}$ of a module $\cal V$.

For a highest weight module with highest weight $(j^{R},k)$, there exists
the following relations
\beqn
\chi_V(z+1,\tau)=S\chi_V(z,\tau)e^{i2\pi j}.
\label{003}
\eeqn
{}From eqs.(\ref{001}, \ref{002}), we also have
\beqn
\chi_V^{R}(z,\tau)=\chi_V^{NS}(-\tau-z,\tau)e^{i\pi k(z+\tau/2)}.
\label{004}
\eeqn
The eqs. (\ref{003}, \ref{004}) make our study of these
(super)characters easier. What we are interested are the modular
transformations of the (super)characters of the admissible representations.
{}From the modular transformation
properties of $S\chi^{R}$, the modular transformation
of the others can be easily obtained. Under the $S$ transformation: $z\rar
z/\tau,
\tau\rar -1/\tau$, we have
\beq
\chi^{R}_{m,s}(z/\tau,-1/\tau)&=&e^{i\pi k(-z-\tau/2)-i2\pi j_{m,s}^{R}}
S_{\td{q}-m,q-s}^{m'\ ,s'} S\chi^{NS}_{m',s'}(z,\tau);\no\\
S\chi^{R}_{m,s}(z/\tau,-1/\tau)&=&
S_{m\ ,s}^{m',s'} S\chi^{R}_{m',s'}(z,\tau);\no\\
S\chi^{NS}_{m,s}(z/\tau,-1/\tau)&=&e^{i\pi k/\tau(z-1/2)-i2\pi j_{m,s}^{R}}
S_{m\ ,s}^{m',s'} \chi^{R}_{m',s'}(z,\tau);\\
\chi^{NS}_{m,s}(z/\tau,-1/\tau)&=&e^{-i\pi k(z+\tau/2+1/(2\tau)-z/\tau)+
i2\pi j_{m,s}^{R}+i2\pi j_{m',s'}^{R}}\no\\
&&S_{m\ ,s}^{m',s'} \chi^{NS}_{m',s'}(z,\tau);\no
\eeq
where $S^{m',s'}_{m\ ,s\ }$ is given as in eq.(\ref{9331910}).
Under $T$ transformation: $\tau\rar\tau+1$,
\beq
\chi_V^{R}(z,\tau+1)&=&e^{i2\pi h_j^{R}}\chi_V^{R}(z,\tau);\no\\
S\chi_V^{R}(z,\tau+1)&=&e^{i2\pi h_j^{R}}S\chi_V^{R}(z,\tau);\no\\
\chi_V^{NS}(z,\tau+1)&=&e^{i2\pi h_j^{NS}}S\chi_V^{NS}(z,\tau);\\
S\chi_V^{NS}(z,\tau+1)&=&e^{i2\pi h_j^{NS}}\chi_V^{NS}(z,\tau);\no
\eeq
We see that under $S$ transformation
\[S\chi^{R}\rar S\chi^{R},~~~~\chi^{NS}\rar\chi^{NS},~~~~\chi^{R}
\leftrightarrow S\chi^{NS}, \]
and under $T$ transformation
\[S\chi^{R}\rar S\chi^{R},~~~~\chi^{R}\rar\chi^{R},
{}~~~~\chi^{NS}\leftrightarrow S\chi^{NS}, \]
which is the reminiscence of the free fermion theory
on torus with various boundary
conditions. The correspondence between the (super)characters and the boundary
conditions on torus can be illustrated by fig.\ 1, where P ( A, resp.)
stands for the periodic ( antiperiodic, resp.)
boundary condition.\\
\begin{picture}(450,120)(-15,-20)
\put(-15,50){$S\chi^{R}$:}
\put(100,50){$\chi^{R}$:}
\put(210,50){$S\chi^{NS}$:}
\put(320,50){$\chi^{NS}$:}
\put(20,20){\framebox(60,50)[l]{P}}
\put(130,20){\framebox(60,50)[l]{A}}
\put(240,20){\framebox(60,50)[l]{P}}
\put(350,20){\framebox(60,50)[l]{A}}
\put(20,20){\vector(0,1){50}}
\put(20,20){\vector(1,0){60}}
\put(130,20){\vector(0,1){50}}
\put(130,20){\vector(1,0){60}}
\put(240,20){\vector(0,1){50}}
\put(240,20){\vector(1,0){60}}
\put(350,20){\vector(0,1){50}}
\put(350,20){\vector(1,0){60}}
\put(20,75){$\tau$}
\put(85,20){$\sigma$}
\put(130,75){$\tau$}
\put(195,20){$\sigma$}
\put(240,75){$\tau$}
\put(305,20){$\sigma$}
\put(350,75){$\tau$}
\put(415,20){$\sigma$}
\put(50,20){\makebox(0,0)[b]{P}}
\put(160,20){\makebox(0,0)[b]{P}}
\put(270,20){\makebox(0,0)[b]{A}}
\put(380,20){\makebox(0,0)[b]{A}}
\put(200,0){\makebox(0,0)[bc]{Figure 1: Boundary conditions on the torus}}
\end{picture}

\section{Structure of the Verma Module}
\label{sec-Structure}
In section 3, we have formulated the characters via coset
construction for the admissible
representations of $\os$, from which we conjecture intuitively the
structure of the corresponding  Verma module.

In this section, we study the Verma module over a contragradient superalgebra
via a quite different approach. The structure of the Verma module  is analysed
by generalizing the Kac-Kazhdan formula \cite{KK} to the case of superalgebra.
Applying this result to $\os$, we reproduce the
(super)character in section 4 for the admissible representation. The  two
important tools, the Jantzen filtration  and the Casimir operator, are
carried over to the super case.
Finally, we give a explicit form for the construction of the singular vectors
in the Verma module of $\os$, generalizing MFF's result in ref.\cite{MFF}.
\subsection{Structure of Verma module}
Let us list some notations \cite{K2,LZ1} useful for our later discussion.\\
Notation:\\
$H$:the abelian diagonalizable subalgebra;\\
$\Delta $ : the set of all roots;\\
$\Delta  ^{+}$ : the set of all positive roots;\\
$\Delta  _{\bar{0}}$ : the set of all even roots;\\
$\Delta  _{\bar{1}}$ : the set of all odd roots;\\
$\Delta  ^{+}_{\bar{0}}$ : $\Delta ^{+}\cap \Delta _{\bar{0}}$;\\
$\Delta  ^{+}_{\bar{1}}$ : $\Delta ^{+}\cap \Delta _{\bar{1}}$;\\
${\cal G}_{\af}$: root space with root $\af$;\\
$(\ ,\ )$ : the nondegenerate symmetric bilinear form on ${\cal G}$,
$(a,b)=(-1)^{deg(a)deg(b)}(b,a)$;\\
$e^{i}_{\af}$: basis of ${\cal G}_{\af}, i=1,\ldots,dim{\cal G}_{\af}$,
such that $(e_{\af}^{i},e_{\bt}^{j})=\delta _{\af +\bt,0}\delta _{i,j}$,
for $\af\in\Delta_+$;\\
$h_\af$ : $\in H,~[e_\af,e_{-\af}]=(e_\af,e_{-\af})h_\af$;\\
$F(\ ,\ )$: the symmetric bilinear form on  $\cal{U(G)}$;\\
$F_{\eta}(\ ,\ )$: the matrix of $F(\ ,\ )$ restricted to ${\cal U(G)}_{-\eta}
\otimes {\cal U(G)}_{\eta}$ with a basis of ${\cal U(G)},
\eta \in \Gamma^{+}$;\\
$M(\lambda),W(\lambda),L(\lambda)$: Verma module, Wakimoto module,and the
irreducible module with HW $\lambda$, the definition see, for example
\cite{K3,Wo,FFr,LZ1};\\
$\rho: \in H^{*}$ such that $\rho(h_{i})=1$.\\
We introduce a Casimir operator on a ${\cal G}$-module.
\blm
Let $V$ be a ${\cal G}$-module, $\Omega$ is an operator on $V$,
\beqn
\Omega (v)=(\mu +2\rho,\mu)+2 \sum_{\af \in \Delta ^{+}}\sum_{i}
e^{(i)}_{-\af}e^{(i)}_{\af}(v),~\forall v \in V_{\mu},\eeqn
 then
(a).\beqn  [\Omega,g]=0, \forall g \in {\cal G};\eeqn
 (b). if $V=M(\lambda),~W(\lambda) \mbox{ or } L(\lambda)$,
\beqn
\Omega =(\lambda+2\rho, \lambda ) {\bf 1}.
\eeqn
\elm
Proof. See proposition 2.7 in ref.\cite{K2} ;
it follows from direct computation.

For an affine Kac-Moody superalgebra, up to a constant, $\Omega =(L_{0}+d)
(k+\bar{g})$, where $L_{0}$ is the zero-mode of the Virasoro algebra
constructed by the Sugawara construction. Sometimes we might use $L_0$
instead of $-d$, when no confusions are made.

Since in a representation $\pi$ of ${\cal G}$ , up to a constant, $\pi(e_
{2\af})=\pi(e_{\af})^{2}$, when $\af \in \Delta _{\bar{1}}$, we can identify
$(e_{\af})^{2}$ with $e_{2\af}$ in $\cal {U(G)}$.
So we select basis of  ${\cal U(G)}_{-}$, which take the following
form
\beq
e^{n_{\af_{1}}}_{\af_{1}}e^{n_{\af_{2}}}_{\af_{2}}\ldots
 e^{n_{\af_{k}}}_{\af_{k}}
   \label{6.1}
\eeq
where $n_{\af_{i}}=0,1$, for $\af \in \Delta _{\bar{1}}$,
$n_{\af_{i}}\in Z_{+}$, for $\af \in \Delta _{\bar{0}}$.

{\bf Definition}: for $\eta \in \Gamma_{+}$, a partition of $\eta$
is a set of non-negative integers  $\{ n_{\af_{i}} \}$,
where $n_{\af_{i}}=0,1$, for $\af_i \in \Delta _{\bar{1}}$,
$n_{\af_{i}}\in Z_{+}$, for $\af_i \in \Delta _{\bar{0}}$, which satisfy
\beq
\sum_{\af_{i}}n_{\af_{i}}\af_{i}=\eta.
\eeq
{\bf Definition}: partition function $P(\eta)$ is the number of all
partitions of $\eta$.\\
 Now consider the leading term of
$detF(\eta)$, by which we mean the monomial term in $detF(\eta)$ with
the maximal power of $h_\af$'s. We get the following lemma which is proven
in the appendix A.
\blm\label{lm02}
(c.f. \cite{KK}, Lemma 3.1) Let $\eta \in \Gamma_+$, then up to a constant
factor, the leading term of det$F_\eta$ is
\beqn
\prod_{\af \in \Delta^+}\prod_{n=1}h_{\af}^{P(\eta-n\af)}
/\prod_{\af \in \Delta_1^+}\prod_{n=1}h_{\af}^{2P(\eta-2n\af)}\no\\
=\prod_{\tiny\begin{array}{c}\vspace{-1ex}\af \in \Delta^+_{\bar{0}} \\
\vspace{-1ex} \af /2
\not \in
\Delta^+_{\bar{1}} \end{array}}
\prod_{n=1}^\infty h_\af^{P(\eta-n\af)}
\prod_{\af \in \Delta^+_{\bar{1}}}
\prod_{\tiny\begin{array}{c}\vspace{-1ex}n=1,\\
\vspace{-1ex} n\in\mbox{odd}\end{array}}
^\infty h_\af^{P(\eta-n\af)},
\eeqn
where the roots are taken with their multiplicities.
\elm
So far we have formulated the Casimir operator and the leading
 term in det$F_\eta$. Then in the same consideration as over Lie algebra
\cite{KK}, we get
\btm
(c.f.\cite{KK}, Theorem 3.1) Generalized Kac-Kazhdan Formula:\\
Let $\cal G=G(A)$ be a contragradient superalgebra with a Cartan matrix
A defined in section 2.1. Then  up to non-zero constant factor
\beq
detF_\eta&=&\prod_{\af \in \Delta^+}\prod_{n=1}\Phi_n(\af)^{P(\eta-n\af)}
/\prod_{\af \in \Delta_1^+}\prod_{n=1}\Phi_{2n}(\af)^{2P(\eta-2n\af)}\no\\
&=&\prod_{\tiny\begin{array}{c}\vspace{-1ex}
\af \in \Delta^+_{\bar{0}}\\ \vspace{-1ex}\af/2\not\in \Delta^+_{\bar{0}}
\end{array}}
\prod_{n=1}^\infty \Phi_n(\af)^{P(\eta-n\af)}
\prod_{\af \in \Delta^+_{\bar{1}}}
\prod_{\tiny\begin{array}{l}\vspace{-1ex} n=1,\\ \vspace{-1ex} n
\in \mbox{ odd}\end{array}}
^\infty \Phi_n(\af)^{P(\eta-n\af)}\label{5.100}
\eeq
where
\beqn
\Phi_n(\af)=h_\af+\rho(h_\af)-n/2(\af,\af),
\eeqn
and the roots are taken with their multiplicities.
\etm
Proof is parallel to that of
Kac-Kazhdan \cite{KK}, by using lemmas 1, 2 and the Jantzen
filtration.

\subsection{Verma module over $\os$}
In this subsection we shall apply the results of the last subsection
to a particular case, namely, the $\os$ superalgebra. The results obtained
in section 3 by the decomposition of $\os$ module are rederived, now, from
the pure algebraic relations. Further more, all the
$\os$ Verma modules are completely classified.

The Chevalley basises of $\os$ are
\beqn\begin{array}{lll}
e_0=\sqrt{2}j_0^{\frac{1}{2}},&f_0=\sqrt{2}j_0^{-\frac{1}{2}},&h_0=4J_0^3 \\
e_1=J_{1}^{-},&f_1=J_{-1}^{+},&h_1=-2J_0^3+k
\end{array}\label{9332401}\eeqn
The Cartan matrix is
\beqn
A=\left( \begin{array}{cc} 2 &-4\\ -1 & 2 \end{array} \right)
\eeqn
and $ e_0, f_0 \in G_{\bar{1}}$,
with the Abelian sub-algebra $H=\{ h_{0}, h_{1}, d\}$.
As in the Kac-Moody Lie algebra \cite{K2,K3}, we can define
a bilinear invariant form
in $\os$ and $H^*$.
Rewrite $A$ as
\beqn
A=
\left(
\begin{array}{cc} 4&0\\0&1\end{array}
\right) \left(
\begin{array}{cc} 1/2&-1\\-1&2\end{array}
\right)
=\mbox{diag}(\epsilon_1,~\epsilon_2)\left( h_{ij} \right)
.\eeqn
The bilinear form on $H$,
\beqn
(h_0,h_0)=8,\ (h_1,h_1)=2,\ (h_0,h_1)=-4, (d, h_i)=(d, d)=0
\eeqn
On $H^*$,
\beqn
(\alpha_0,\alpha_0)=1/2,\ (\alpha_1,\alpha_1)=2,\ (\alpha_0,\alpha_1)=-1.
\eeqn
The fundamental dominant weights $\Lambda_0,~\Lambda_1$ satisfy
\beqn
\Lambda_i(h_j)=\delta_{ij}, \ \Lambda_i(d)=0
\eeqn
We see that $\Lambda_1=2\Lambda_0+\alpha_1/2$ and
\beqn
(\Lambda_0, \Lambda_0)=0, \ (\alpha_0,\Lambda_0)=1/4,\ (\alpha_1,\Lambda_1)=1
\eeqn
\beqn
\rho=\Lambda_0+\Lambda_1
\eeqn
\beq
\Delta^+:&\{\alpha=n_0\alpha_0+n_1\alpha_1,\ \ \
2n_1-n_0=\pm 2,\pm 1, 0,\ \ \
n_1\geq 1~~\mbox{or } n_1=0,\ \ \  n_0=1,2\} \no\\
\Delta_1^+:&\{\alpha=n_0\alpha_0+n_1\alpha_1,\ \ \ 2n_1-n_0=\pm 1,\ \ \
n_1\geq 1~~\mbox{or } n_1=0, n_0=1\}
\eeq
A highest weight with level $k$, isospin $j$, is
\beq
\Lambda =4j\Lambda_0+(k-2j)\Lambda_1
\eeq
{}From
\beq
(\Lambda+\rho,\alpha)&=&\frac{1}{4}(n_0-2n_1)(4j+1)+(k+3/2)n_1, \no\\
(\alpha,\alpha)&=&\frac{1}{2}(n_0-2n_1)^2,
\eeq
we have
\beq
\Phi_{\alpha,n}&=&(\lambda+\rho,\alpha)-\frac{n}{2}(\alpha,\alpha) \no\\
&=&\frac{1}{4}(n_0-2n_1)(4j+1)+(k+3/2)n_1-\frac{n}{2}(n_0-2n_1)^2
\eeq
After direct computation, we get
\blm \label{lm05}
$M_j$ is reducible if and only if
\beq
4j+1=m-s(2k+3), \ \ \mbox{assuming } 2k+3 \neq 0,\label{5.10}
\eeq
for some $m,s\in Z~~, m+s\in \mbox{ odd }, m<0,s<0$ or $m>0, s\geq 0$.
if eq.(\ref{5.10}) holds,
then there exists a singular vector in $M_j$ with isospin
$j_{-m,s}$.
\elm
\btm (c.f. \cite{FF} Theorem 4.1)
Let $k+3/2\neq 0$ then the structure of Verma module $M(j)$ is described
by one of the following diagram.
\beq
\begin{array}{cccccccccc}
v_0&v_0&&v_0&&&v_0&&v_0&v_0\\
\cdot&\downarrow&\swarrow&&\searrow&
\swarrow&&\searrow&\downarrow&\downarrow \\
&v_1&v_1&&v_{-1}&v_1&&v_{-1}&
v_1&v_1\\
&&\downarrow&\large{\crossst}&\downarrow
&\downarrow&\large{\crossst}&\downarrow
&\downarrow&\downarrow\\
&&v_2&&v_{-2}&v_2&&v_{-2}&v_2&v_2\\
&&\downarrow&\large{\crossst}&\downarrow
&\downarrow&\large{\crossst}&\downarrow
&\downarrow&\downarrow\\
&&v_3&&v_{-3}&v_3&&v_{-3}&v_3&v_3\\
&&\vdots&&\vdots
&\vdots&\vdots&\vdots
&&\vdots\\
&&v_{i-1}&&v_{1-i}&v_{i-1}&&v_{1-i}&v_{i-1}&v_{i-1}\\
&&\downarrow&\large{\crossst}&\downarrow&\searrow&&\swarrow
&\downarrow&\downarrow\\
&&\vdots&\vdots&\vdots&&v_i&&\vdots&v_i\\
I&I\!I&&I\!I\!I_-&&&I\!I\!I_+&
&I\!I\!I_-^0&I\!I\!I_+^0
\end{array}
\eeq
where $v_i$'s are the singular vectors in M(j). An arrow or a chain
of arrows, goes from a  vector to another iff the second vector is in the
sub-module generated by the first one.
\etm
Proof follows from lemma \ref{lm05} and the generalized Kac-Kazhdan
formula eq.(\ref{5.100}). It is analogous to that in ref.\cite{BF,FF}.

If $k  \notin Q$, then case $I$ and case $I\!I$ occurs.
If $k\in Q$, we write $2k+3=\tilde{q}/q$,
where $\td{q}+q\in \mbox{ even, }gcd(\tilde{q},(q+\td{q})/2)=1$.
If eq.(\ref{5.10}) holds, then for
 $2k+3>0(<0)$, case $I\!I\!I_-~~(I\!I\!I_+)$ occurs.
If $j=j_{0,s}$ for some $s\in odd$, then $I\!I\!I_-^0$ or $I\!I\!I_+^0$ occurs.
Note that
\beq
\Phi_{\alpha,n}(\lambda)=-\Phi_{\alpha,n}(-2\rho-\lambda+n\alpha).
\eeq
\bpp
A singular vector $v_1$ with weight $\lambda_1$ is in a Verma module
generated by $v_2$ with weight $\lambda_2$ iff a singular vector $v_2'$
with weight $\lambda_2'=-2\rho-\lambda_2$ is in the Verma module generated
by $ v_1'$ with weight $\lambda_1'=-2\rho-\lambda_1$.
\epp
{\it Remark}: The above duality relation is similar to that of Virasoro
algebra, where
we have the following duality, $(h, c)\leftrightarrow (1-h, 26-c)$.

We note that case $I\!I\!I_-, I\!I\!I_+ $ can be described by a subdiagram of
diagram 1
and diagram 2 (resp.)
\beq
\begin{array}{cccccccccccc}
v_{m,s}&\rightarrow&v_{-m+2\tilde{q},s}&\rightarrow&
v_{m+2\tilde{q},s}&\rightarrow&\cdots&\rightarrow&
v_{m+2n\tilde{q},s}&\rightarrow &v_{-m+2(n+1)\tilde{q},s}&\rightarrow\\
&\searrow&&\large{\crossrd}&&\large{\crossrd}&&\large{\crossrd}&&\large{\crossrd}&&\\
&&v_{-m,s}&\rightarrow&v_{m-2\tilde{q},s}&\rightarrow
&\cdots&\rightarrow&v_{m-2n\tilde{q},s}&\rightarrow&
v_{-m-2n\tilde{q},s}&\rightarrow\\
v_{0,s}&\rightarrow&
v_{2\tilde{q},s}&\rightarrow&
v_{-2\tilde{q},s}&\rightarrow&
\cdots&\rightarrow&v_{2n\tilde{q},s}&\rightarrow&
v_{-2n\tilde{q},s}&\rightarrow\\
v_{\tilde{q},s}&\rightarrow&
v_{-\tilde{q},s}&\rightarrow&
v_{-3\tilde{q},s}&\rightarrow&
\cdots&\rightarrow&v_{(2n+1)\tilde{q},s}&\rightarrow&
v_{-(2n+1)\tilde{q},s}&\rightarrow\\
\end{array}\eeq
\begin{center}Diagram. 1\end{center}
\beq
\begin{array}{cccccccccccc}
v_{-m,s}&\leftarrow&v_{m-2\tilde{q},s}&\leftarrow&
v_{-m-2\tilde{q},s}&\leftarrow&\cdots&\leftarrow&
v_{-m-2n\tilde{q},s}&\leftarrow &v_{m-2(n+1)\tilde{q},s}&\leftarrow\\
&\nwarrow&&\large{\crossth}&&\large{\crossth}&&\large{\crossth}&&\large{\crossth}&&\\
&&v_{m,s}&\rightarrow&v_{-m+2\tilde{q},s}&
\leftarrow&\cdots&\leftarrow&v_{-m+2n\tilde{q},s}&\leftarrow&
v_{m+2n\tilde{q},s}&\leftarrow\\
v_{0,s}&\leftarrow&
v_{-2\tilde{q},s}&\leftarrow&
v_{2\tilde{q},s}&\leftarrow&
\cdots&\leftarrow&v_{2n\td{q},s}&\leftarrow&v_{-2(n+1)\td{q},s}&\leftarrow\\
v_{-\tilde{q},s}&\leftarrow&
v_{\tilde{q},s}&\leftarrow&
v_{-3\tilde{q},s}&\leftarrow&
\cdots&\leftarrow&v_{-(2n+1)\td{q},s}&\leftarrow&v_{(2n+1)\td{q},s}&\leftarrow\\
\end{array}
\eeq
\begin{center}Diagram. 2\end{center}
Here $0<  m<\td{q},0\leq s <q,$ and the sum of the two subscripts is always
odd. From the above theorem, we get the following corollary about the
relation between the irreducible module and the Verma module.
\bc \label{c01}
Let $k+3/2\neq 0$, for any irreducible module $L_{v_0}$, there exists
a sequence which is a resolution of $L_{v_0}$,
\beqn
\cdots \stackrel{\partial_{i-1}}{\rar} M^{i}
\stackrel{\partial_i}{\rar}
M^{i+1}\stackrel{\partial_{i+1}}{\rar} \cdots
\stackrel{\partial_{-2}}{\rar}
M^{-1} \stackrel{\partial_{-1}}{\rar} M^0=M_{v_0}\rar 0,
\eeqn
where $M^i$ is direct sum of Verma modules.
\ec
Proof. (i) If $M_{v_0}$ is in the case (I), then simply let $M^i=0$,
for $i<0$.\\
($i\!i$) If $M_{v_0}$ is in the case ($I\!I$) or ($I\!I\!I^0_{\pm}$),
let $M^i=0$, for $i<-1$; $M^{-1}=M_{v_1}$, $\df_{-1}$ is an embedding.\\
($i\!i\!i$) If $M_{v_0}$ is in the case ($I\!I\!I_-$), let $M^i=
M_{v_i}\oplus M_{v_{-i}}$,for $i<0$, $\df_{i}: (x,y)\rar (x-y,x-y)$,
for $i<-1$;$\df_{-1}: (x,y)\rar x-y$. \\
($iv$) If $M_{v_0}$ is in the case ($I\!I\!I_+$), let $M^i=
M_{v_i}\oplus M_{v_{-i}}$, for $-n<i<0$
$M^{-n}=M_{v_n}.\  \df_{i}: (x,y)\rar (x-y,x-y)$, for $-n<i<-1$
$\df_{-1}: (x,y)\rar x-y,\ \df_{-n}: x\rar(x,x)$.\\
Then the corollary is easily verified.

\subsection{MFF Constrution}
We take steps after ref.\cite{MFF}, where a singular vector in a Verma module
over $\widehat{SL(2)}$ is given  explicitly. The crucial point is that
the authors of \cite{MFF}
generalized polynomially commutators between generators in $\cal U(G)$
 to commutators between those with complex exponents. Similarly we also can
generalize
this procedure to contragradient \lsa. However we only discuss it on $\os$.

Let $M_{j_{m,s}}$ be a Verma module as that in the lemma \ref{lm05}, then
\begin{enumerate}
\item if $m>0,\ s\geq 0$
\beqn
|-m,s\rangle=f_0^{m+s\td{q}/q}f_1^{1/2[m+(s-1)\td{q}/q]}f_0^{m+(s-2)\td{q}/q}
\cdots f_1^{1/2[m-(s-1)\td{q}/q]}f_0^{m-s\td{q}/q}|m,s\rangle\label{9331920}
\eeqn
\item if $m<0,\ s<0$
\beq
|-m,s\rangle&=&f_1^{-1/2[m+(s+1)\td{q}/q]}f_0^{-[m+(s+2)\td{q}/q]}
f_1^{-1/2[m+(s+3)\td{q}/q]}
\cdots\no\\
&&\cdots f_0^{-[m-(s+2)\td{q}/q]}f_1^{-1/2[m-(s+1)\td{q}/q]}|m,s\rangle
\label{9331921}
\eeq
\end{enumerate}
is a null vector. The action of a multiplying factor in the r.\ h.\ s.\
of the eqs.(\ref{9331920},\ref{9331921}) on the weight space
is as the corresponding fundamental Weyl reflection.
We see that $m+s \in \mbox{odd},\ m,\ s<0,\mbox{ or }m>0,\ s\geq 0$
is equivalent to the following restraints,
\begin{enumerate}
\item the sum of all the exponents of a fixed Chevalley generator
is an non-negative integer.
\item the sum of all the exponents of a fixed odd Chevalley generator
plus the times it appears in the expression is odd.
\end{enumerate}
The second restraint is of great importance for the contragradient
superalgebras. More exactly, we should rewrite the multiplying factor
$f_0^x$ as $f_0(f_0^2)^{(x-1)/2}$ in the r.\ h.\ s.\ of the above equations.
Define operator
\beqn
F(m,s,t)=f_0(f_0^2)^{[m+st-1]/2}f_1^{[m+(s-1)t]/2}f_0(f_0^2)^{[m+(s-2)t-1]/2}
\cdots f_1^{[m-(s-1)t]/2}f_0(f_0^2)^{[m-st-1]/2},\label{93381}
\eeqn
where $m>0,\ s\geq 0,\ m+s \in odd,\ t=\td{q}/q$. We assume that $(f_0^2)
^x$ is always an even operator. The sum of all the exponents of $f_0^2$ is
$(m-1)(s+1)/2$, a non-negative integer.
To prove that $|-m,s\rangle$ in eq.(\ref{9331920}) is a singular vector
in the Verma module, firstly, we have to show that that $F(m,s,t)$ defined
in eq.(\ref{93381}) is in the enveloping algebra. To do that we introduce
the following generalized commutators
\cite{MFF},
\beq
[g_1,g_2^\r]&=&\sum_{i=1}^{\infty} \left( \begin{array}{c} \r\\ i\end{array}
\right) g_2^{\r-i}[\cdots[[g_1,\underbrace{g_2],g_2],\cdots g_2}_i]\no\\
&=& -\sum_{i=1}^{\infty}(-1)^i  \left( \begin{array}{c} \r\\ i\end{array}
\right) [\cdots[[g_1,\underbrace{g_2],g_2],\cdots g_2}_i]g_2^{\r-i},~~~g_2\in
{\cal G}_0, \label{1r}\\
{[}g_1^{\r_1},g_2^{\r_2}]&=&\sum_{j_1=1}^{\infty}\sum_{j_2=1}^{\infty}
\left(\begin{array}{c} \r_1\\ j_1\end{array}
\right)\left(\begin{array}{c} \r_2\\ j_2\end{array}
\right)Q_{j_1,j_2}(g_1,g_2)g_2^{\r_-j_2}g_1^{\r_1-j_1},~~~g_1,\ g_2\in
{\cal G}_0, \label{r1r2}
\eeq
where $Q_{j_1,j_2}$'s$\in {\cal U(G)}$ are independent of  $\r_1,\ \r_2$ and
given by induction,
\beq
Q_{j_1,j_2}(g_1,g_2)&=&[g_1,Q_{j_1-1,j_2}]+\no\\
 &&+\sum_{i=0}^{j_2-1}(-1)^{j_2-i-1}\left(\begin{array}{c} j_2\\ i\end{array}
\right) Q_{j_1-1,i}[\cdots[[g_1,\underbrace{g_2],g_2],\cdots g_2}_{j_2-i}],
\eeq
and $Q_{0,0}=1,\ Q_{0,v}=0,\forall v>0$.
By repeatedly using eq.(\ref{1r}) and (\ref{r1r2}), we can rewrite $F(m,s,t)$
in the following form,
\beq
F(m,s,t)=\sum_{j_0=1}^{\infty}\sum_{j_1=1}^{\infty}P_{j_0,j_1}(f_0,f_1)
f_0^{m(s+1)-j_0}f_1^{ms/2-j_1},
\eeq
where $P_{j_0,j_1}\in {\cal U([G_-,G}_{0,-}])$, depends on $t$ polynomially,
with
the power of $t$ less than $j_0+j_1+1$ and that of
$f_0\ (f_1)$ equal to $j_0\ (j_1)$. We see that
$F(m,s,t)$ is in the enveloping algebra $\cal U(G)$ if and only if
\beqn
P_{j_0,j_1}=0,~~~~\mbox{ for }j_0>m(s+1),\mbox{  or }
j_1>ms/2.
\label{93391}
\eeqn
To see that eq.(\ref{93391}) is indeed satisfied, let us check the cases for
which  $t\in \mbox{ odd}$. In that case, each exponent in the r.h.s. of the
eq.(\ref{93381})
is an integer. By the analogy of the discussion
in ref.\ \cite{MFF}, it can be shown that
$F(m,s,t)$ is well defined.  $F(m,s,t)$ is in the enveloping algebra
${\cal U(G)}$, and
gives rise to the singular vector in the Verma module
$M(j_{m,s})$ with isospin $j_{-m,s}$ and level  $k=(t-3)/2$ when acting on
the HWS $|m,s\rangle$.
Since eq.(\ref{93391}) valids for infinitely many $t$'s, i.e.
for those $t\in\ odd$,
it can be deduced
that so does it for all $t\in {\cal C}$, noting that $P_{j_0,j_1}$'s
depend on $t$ polynomially.
Now we come to the conclusion that $F(m,s,t)$ is in the enveloping algebra
${\cal U(G)}$ for all $t \in {\cal C}$.
Secondly, it remains to check that
\beq
e_iF(m,s,t)|m,s\rangle=0,~~~~i=0,\ 1, \label{e0e1}
\eeq
which is equivalent to say that $F(m,s,t)|m,s\rangle$ is singular
vector in $M(j_{m,s})$ with isospin $j_{-m,s}$. Eq.(\ref{e0e1}) can be
verified by contracting $e_i$'s on the r.h.s. of eq.(\ref{9331920}).
To make our discussion more concrete, in appendix B an example
MFF construction is given for $m=2,\ s=1$.

So far we have illustrated that the r.\ h.\ s.\ of eq.(\ref{9331920})
is well defined, the similar discussion can be taken over to the
eq.(\ref{9331921}) for $m,\ s<0,\
m+s\in $ odd.
\section{Wakimoto Module Over $\os$}
In this section we study the Wakimoto Module for $\os$ in the free field
representation \cite{BO1}. Again we obtain results analogous to that
of $\widehat{SL(2)}$ \cite{BF}. The admissible
representation and the Wakimoto module are related by the Felder \cite{Fl,BF}
BRST operator. More exactly the admissible representation is the  zero degree
cohomology of the Felder  complex while other degree cohomology vanish.
The crucial points for the BRST operator are the screening operator
\cite{BF,BO1} (or
interwining operator in \cite{FF} ) and again the Jantzen filtration \cite{J}.

Wakimoto module $W(\lambda)$  over $\widehat{SL(2)}$ was first studied in
ref.\cite{Wo}, later over affine Kac-Moody algebra generally in
ref.\cite{FFr,FF}.
It admits a free field realization \cite{FF}. For superalgebra $\os$,
we also have the free field representation \cite{BO1}.
\beqn \left\{\begin{array}{l}
J^+=-\bt;\\
J^-=\bt \r^2-i\af_+\r \df \phi+\r \psi\psi^+ -k\df \r+(k+1)\psi\df\psi;\\
J^3=-\bt\r+i\af_+/2\df \phi-\frac{1}{2}\psi\psi^+;\\
j^-=\r(\psi^+-\bt\psi)+i\af_+\df\phi+(2k+1)\df \psi;\\
j^+=\psi^+-\bt\psi,\end{array} \right. \label{5.500}
\eeqn
where $\af_+=\sqrt{2k+3}$,
$(\bt,\r)$ are bosonic fields with conformal isospin $(1,0)$, $(\psi^+,\psi)$
are fermionic fields with conformal isospin $(1,0)$.
More concretely, expand them in
Laurent power series:
\beqn \begin{array}{ll}
\bt(z)=\sum_n \bt_n/z^{n+1};&\r(z)=\sum_n \r_n/z^{n};\\
\psi^+(z)=\sum_n \psi^{+}_n/z^{n+1};&\psi(z)=\sum_n \psi_n/z^{n+1};\\
i\df\phi(z)=\sum_n \phi_n/z^{n+1}.& \end{array}
\eeqn
\beqn\begin{array}{ll}
\bt(z_1)\r_(z_2)=1/z_{12};&\psi^+(z_1)\psi(z_2)=1/z_{12};\\
\phi(z_1)\phi_(z_2)=-\ln z_{12},&\end{array}
\eeqn
which is equivalent to the commutators
\beqn\begin{array}{ll}
[\bt_n,\r_m]=\delta_{n+m,0};&\{ \psi^+_n,\psi_m \}=\delta_{n+m,0};\\
{[}\phi_n,\phi_m]=n\delta_{n+m,0},\end{array}
\eeqn
while other commutators vanish.

The HWS in Wakimoto module is annihilated by all positive modes of these field
as well as $\bt_0,\psi^+_0$, and is an eigenstate of $\phi_0$,
\beq
\phi_0|HWS\rangle=2j/\af_+|HWS\rangle.
\eeq
The Wakimoto module is the Fock space  generated by $\bt_n,\r_n,
\psi_n,\psi^+_n,n<0$, and $\r_0,\psi_0$.
 \bpp \label{pp08}
 Let \beq V(z)=\sum_n V_n/z^{n+1}=(\psi^++\bt\psi)e^{i\af_-\phi}(z)&
,~~~\af_-=-1/\af_+, \eeq
($i$) $V(z)$ is a screening operator. i.e. the OPE of the $\os$ currents
and $V(z)$
are total derivatives \cite{BO1}.
($i\!i$) $V(z)$ induces an $\os$-module homomorphism $W_j\rightarrow
W_{j-1/2}$.
\epp
Proof.
($i$) follows from direct computation.\\
($i\!i$) Note that the Fock space is characterized by the fact that it is an
eigenspace of $\phi_0$ with eigenvalue $2j/\af_+$, i.e.
\beq
\forall |v\rangle \in F_j,~\phi_0|v\rangle=2j/\af_+|v\rangle.
\eeq
Since
\beqn
i\df\phi(z_1)V(z_2)=-1/\af_+\frac{V(z_2)}{z_{12}},~~~~
{[}\phi_0,V_n]=-1/\af_+.
\eeqn
{}From ($i$) we have \beq
[J_n^a,V_0]=0.
\eeq
So we can deduce that $V_0: F_j\rightarrow F_{j-1/2}$ is a homomorphism.

Define \beq Q_m=\oint\oint\ldots \oint V(z_1)V(z_2)\ldots V(z_m). \eeq
Then $Q_m$ is an $\os$-module homomorphism,
\beq
Q_m: F_j\rightarrow F_{j-1/2}.
\eeq
The proof is as in ref.\cite{BF,FF} for the
case of $\widehat{SL(2)}$ current algebra.

\bpp
\label{pp10}
Let $2k+3=\td{q}/q\neq 0,\ 4j_{m,s}+1=m-s(2k+3),\ m,s \in Z$, $m+s \in$ odd,
then\\
($i$) if $\td{q}>m>0,\ s\geq 0$, in Wakimoto module $W_{j_{m,s}}$,
there exists one
and only one cosingular vector $|w_{-m,s}\rangle$ with the isospin
$j_{-m,s}$, and under homomorphism map (up to a nonzero constant)
\beqn
Q_m:W_{j_{m,s}}\rar W_{j_{-m,s}},\
Q_m|w_{-m,s}\rangle=|j_{-m,s}\rangle,
\eeqn
where $|j_{-m,s}\rangle$ is the vacuum vector in $W_{j_{-m,s}}$.\\
($i\!i$) if $-\td{q}<m<0,\ s<0$,in Wakimoto module $W_{j_{m,s}}$,there exists
one
and only one singular vector $|w_{-m,s}\rangle$ with the isospin
$j_{-m,s}$, and under homomorphism map(up to a nonzero constant)
\beqn
Q_{-m}:W_{j_{-m,s}}\rar W_{j_{m,s}},\
Q_{-m}|j_{-m,s}\rangle=|w_{-m,s}\rangle,
\eeqn
where $|j_{-m,s}\rangle$ is the vacuum vector in $W_{j_{-m,s}}$.\\
\epp
Proof see appendix C.

By the above proposition, we can work out the weights of singular
vectors and cosingular vectors in a Wakimoto module $W_j$. By using the
techniques such as the Jantzen filtration as those used in studying the
structure of Wakimoto modules over $\widehat{SL(2)}$ \cite{BF,FF}
and Feigin-Fuchs modules
over Vir. \cite{FFu}, we reach the following theorem.
\btm \label{tm04}
(cf.\cite{FF},theorem 4.2) Let $k+3/2\neq 0$, then
  the structure of Wakimoto module $W_j$ can be described
by one of the following diagram.
\beq
&&\begin{array}{ccccccc}
w_0&w_0&w_0&w_0&w_0&w_0&w_0\\
\cdot&\downarrow&\uparrow
&\downarrow&\uparrow&\downarrow&\uparrow\\
&w_1&w_1
&w_1&w_1&w_1&w_1\\
&&
&\uparrow&\downarrow&\uparrow&\downarrow\\
&&
&w_2&w_2&w_2&w_2\\
&&
&\downarrow&\uparrow&\downarrow&\uparrow\\
&&
&w_3&w_3&w_3&w_3\\
&&
&\vdots&\vdots&\vdots&\vdots\\
&&
&\uparrow&\downarrow&\uparrow&\downarrow\\
&&
&w_{n-1}&w_{n-1}&w_{n-1}&w_{n-1}\\
&&
&\vdots&\vdots&\downarrow&\uparrow\\
&&
&&&w_n&w_n\\
I&I\!I(+)&I\!I(-)
&I\!I\!I^0_-(+)&I\!I\!I^0_-(-)&I\!I\!I^0_+(+)&I\!I\!I^0_+(-)
\end{array}\no\\
&&\begin{array}{ccccccccc}
&w_0&&&w_0&&&w_0&\\
\swarrow&&
\nwarrow&\swarrow&&\searrow&\nearrow&&\nwarrow \\
w_1&&w_{-1}&w_1&&
w_{-1}&w_1&&w_{-1}\\
\uparrow&\large{\crossth}&\downarrow
&\uparrow&\large{\crossnd}&\uparrow
&\downarrow&\large{\crossst}&\downarrow\\
w_2&&w_{-2}&w_2&&w_{-2}&w_2&&w_{-2}\\
\downarrow&\large{\crossth}&\uparrow
&\downarrow&\large{\crossst}&\downarrow&\uparrow
&\large{\crossnd}&\uparrow\\
w_3&&w_{-3}&w_3&&w_{-3}&w_3&&w_{-3}\\
\vdots&\vdots&\vdots&\vdots&\vdots&\vdots&\vdots&\vdots&\vdots\\
\uparrow&\large{\crossth}&\downarrow&
\uparrow&\large{\crossnd}&\uparrow
&\downarrow&\large{\crossst}&\downarrow\\
w_{n-1}&&w_{1-n}&w_{n-1}&&w_{1-n}&w_{n-1}&&w_{1-n}\\
\vdots&\vdots&\vdots&\searrow&&\swarrow&\nwarrow&&\nearrow\\
&&&&w_n&&&w_n&\\
&I\!I\!I_-&&&I\!I\!I_+(+)&
&&I\!I\!I_+(-)&
\end{array}
\eeq
where $w_i$'s are the singular vectors in $W_j$ or in its subquotient
corresponding to $v_i$ in the Verma modules, i.e. they have the same weight.
An arrow or a chain
of arrows, goes from the vector to another iff the second vector is in the
sub-module generated by the first one.
\etm

Having worked out the structure of the Wakimoto modules,
we get the a resolution of the admissible module in terms of Wakimoto
modules.
\bc\label{c02}
The following sequence
 is a resolution of admissible module $L_{m,s}$,
\beqn
\cdots \stackrel{Q_m}{\rar} W_{-m+2n\td{q},s}
\stackrel{Q_{\td{q}-m}}{\rar}
W_{m+2(n-1)\td{q},s}\stackrel{Q_{m}}{\rar} \cdots
\stackrel{Q_{\td{q}-m}}{\rar}
W_{-m+2\td{q},s} \stackrel{Q_{\td{q}-m}}{\rar} W_{m,s}\stackrel{Q_m}{\rar}
W_{-m,s}\stackrel{Q_{\td{q}-m}}{\rar}\cdots,
\eeqn
\ec
The proof of the nilpotency and that the resolution is the irreducible
module is similar to that of ref.\cite{BF} for the admissible modules over
$\widehat{SL(2)}$ and ref.\cite{Fl} for the irreducible modules in
the minimal models over Virasoro algebra.

\section{Conclusions and Conjectures}
In this paper, we have studied the modules over
affine Kac-Moody superalgebras in general.
A naive approach to find the structure of superalgebra module is by coset
construction $G=G_0\otimes G/G_0$, where $G_0$ is generated by the even
generators. Such decomposition is analysed in detail for $\os$, for which
$G_0=\widehat{SL(2)}$ in section 3.
Our conjecture is that for a more general affine Lie
superalgebra, the coset space $G/G_0$ will be generated by the $W_N$
algebra.

Besides the coset decomposition, a more general procedure of classifying
the superalgebra modules is provided by generalizing the Kac-Kazhdan
formula to the super case. Indeed it is possible as we have done it
in section 4. The corresponding MFF construction of null vectors is
generalized in a similar fashion.

In parallel, we have also analysed the Wakimoto
module for the affine superalgebra.
All these results will be relevant for our analysis of the $G/G$ gauged WZNW
model on a supergroup manifold, which is the subject of our forthcoming
paper(\cite{FY}).

\bigskip
\noindent
{\bf Acknowledgement}:
We are grateful to H.Y. Guo, H.L. Hu, K. Wu and R.H. Yue for useful discussions
and suggestions.
This work is supported in part by the National Science Foundation
of China and the National Science Committee of China.

\appendix
\section{The Leading Term of the $Det F_\eta $}
Now we prove the lemma \ref{lm02} in section \ref{sec-Structure}.
First, we prove several propositions.\\
{\bf Definition}: $$P_{\af_{i}}(\eta)=\sum_{\tiny
\begin{array}{c}\vspace{-1ex}\{ n_{\af_{j}}\}
\mbox {partition of } \eta \\
\vspace{-1ex} \af_i\in \{\af_j\}\end{array}}n_{\af_i} $$.
\bpp
\beq
\sum_{\eta \in \Delta_{+}} P(\eta)e^{-\eta}=
\prod_{\af \in\Delta  ^{+}_{\bar{0}}}
\frac{1}{1-e^{-\af}}\prod_{\bt \in \Delta  ^{+}_{\bar{1}}}{(1+e^{-\bt})}.
\eeq
\epp
Proof. It is obtained by  direct computation.
\bpp\label{pp03}
\beq
P_{\af_{i}}(\eta)=\left\{ \begin{array}{l}
\sum_{n=1}P(\eta-n\af_i),
\mbox{ if } \af_{i} \in \Delta _{\bar{0}};\\
\sum_{n=1}P(\eta-n\af_i)(-1)^{n-1},
\mbox{ if }\af_{i} \in \Delta _{\bar{1}}
                          \end{array} \right.
\eeq
\epp
Proof. Using the generating function,
\beq
\sum_{\eta} P_{\af_{i}}(\eta)e^{-\eta}
&=&\sum_\eta
\sum_{\{n_{\af_{j}}\} \mbox{partition of }
\eta}n_{\af_i}exp[-\sum_{\af_j}n_{\af_j}\af_j]\\
&=&\frac{
\prod_{\af_j\neq \af_i,\  \af_j\in\Delta_1^+}(1+e^{-\af_j})}
{\prod_{\af_{j}\neq \af_{i},\
\af_j \in\Delta_0^+}(1-e^{-\af_j})}
\sum_{n_{\af_i}}n_{\af_i}e^{-n_{\af_i}\af_i};\label{9331705}
\eeq
note that
\beqn
\sum_{n_{\af_{i}}}n_{\af_{i}}e^{-n_{\af_i}\af_i}=\left\{
\begin{array}{ll}
\frac{e^{-\af_{i}}}{(1-e^{-\af_{i}})^{2}},
& \mbox{if }\af_j \in \Delta_{\bar{0}}; \\
e^{-\af_{i}},
& \mbox{if }\af_j \in \Delta_{\bar{1}};
\end{array} \right.\label{9331706}
\eeqn
by proposition \ref{pp03}, we have
\beqn
\sum_\eta P_{\af_i}(\eta)e^{-\eta}=\left\{\begin{array}{ll}
\sum_\eta P(\eta)e^{-\eta}
\sum_{n=1}^{\infty}e^{-n\af_i},&\mbox{when } \af_i \in \Delta_0;\\
\sum_\eta P(\eta)e^{-\eta}
\sum_{n=1}^{\infty}e^{-n\af_i}(-1)^{n-1},&\mbox{when } \af_i \in \Delta_0.
\end{array}\right.\label{9331707}
\eeqn
By using eqs.(\ref{9331706},\ref{9331707}) and
comparing the coefficients of the term $e^{-\eta}$ in both sides of
eq.(\ref{9331705}),
we get the proposition.

Now selecting the basis of $\cal{U(G)}_\eta$ as in eq.(\ref{6.1}),
where $\{ n_{\af_i} \}$ is a partition of $\eta$. Note that the leading term is
 the product of the leading terms in the diagonal elements of the matrix
$F_\eta$, which is proportional to
\beq
\prod_{\af_i} h_{\af_i}^{P_{\af_i}(\eta)}.
\eeq
Moreover, note that $h_{2\af_i}=2h_{\af_i}$, when $\af_i \in \Delta_{\bar{1}}$,
which completes the proof of the lemma.
\section{The singular vector $|-2,1\rangle $}
In section 4, it is shown that the explicit form of the singular vector
in a Verma module over $\os$ can be constructed by the use of
eqs.(\ref{9331920},\ref{9331921}).
To elucidate such a procedure, we consider a simple
case, i.e. $m=2,\ s=1$.

By definition,
\beq
F(2,1,t)=f_0(f_0^2)^{\frac{1+t}{2}}f_1f_0(f_0^2)^{\frac{1-t}{2}}.
\label{ap2001}
\eeq
Here, for $\os$ (see eq.(\ref{9332401})),
\beq
&&f_0=\sqrt{2}j^-_0,\ \ \ f_0^2=-2J^-_0,\ \ \ f_1=J^+_{-1}\no\\
&&{[}\cdots[f_1,\underbrace{f_0^2],\cdots,f_0^2}_i]=0,\ \ \ \forall i\geq 3.
\eeq
By using eq.(\ref{1r}), eq.(\ref{ap2001}) can be rewritten as,
\beq
F(2,1,t)=f_0f_1f_0^3-
\frac{t+1}{2}f_0[f_1,f_0^2]f_0+
\frac{t^2-1}{8}[[f_1,f_0^2], f_0^2]
.\eeq
More concretely,
\beq
F(2,1,t)=4J_{-1}^+(J_0^{-})^2-4j_{-1}^+j_0^-J_0^-
-4(t+1)J_{-1}^3J^-_0+2(t+1)j^-_{-1}j^-_0
-(t^2-1)J^-_{-1}
.\eeq
So it is obvious that $F(2,1,t)\in {\cal U(G)}$.
After the direct computation, we can get
\beq
[e_0, F(2,1,t)]&=&(f_1f_0^3-f_0f_1f_0^2)((-2+h_0)\frac{1-t}{2}-\frac{t^2-1}{2})
+(f_0^2f_1f_0-f_0^3f_1)(\frac{t+1}{2}h_0+\frac{t^2-1}{2})\no\\
{[}e_1, F(2,1,t)]&=&f_0^4(-t+2+h_1)
\label{ap003}
.\eeq
For the HWS $|2,1\rangle , \ \ \ t=2k+3$,
\beq
h_0|2,1\rangle =&4J_0^3|2,1\rangle&=(1-t)|2,1\rangle \no\\
h_1|2,1\rangle =&(-2J_0^3+k)|2,1\rangle&=(t-2)|2,1\rangle
\label{ap005}
.\eeq
Combining eqs.(\ref{ap003},\ref{ap005}), we see that
\beq
e_iF(2,1,t)|2,1\rangle =0,\ \ \ i=0,1
.\eeq
So $F(2,1,t)|2,1\rangle $ is a singular vector $|2,1\rangle$
in $M_{j_{2,1}}$.

\section{The Non-vanishing of $Q_m|w_{-m,s}\rangle $
and $Q_{-m}|j_{-m,s}\rangle $ }
In this appendix, we manage to prove the proposition \ref{pp10},
\beq
Q_m|m,-s\rangle\neq 0,~~~~\mbox{ if }0<m<\td{q},~s>0,~s+m\in~\mbox{ odd}.
\eeq
Due to the presence of the fermionic operator in the screening operator $V(z)$,
it takes much effort to complete our proof, in contrast to
the analogous conclusions \cite{Fl,BF} on the
study of the Feigin-Fuchs modules over Virasoro algebra (or Wakimoto
modules over $\widehat{SL(2)}$).

Now consider
\beq
Q_m=\oint d\!z\oint \prod_{i=2}^m d\!z_iV(z)\prod_{i=2}^m V(z_i),
\eeq
where $V(z)=(\psi^++\bt\psi)e^{i\af_-\phi}(z)$, and the integration contour
is depicted in fig.\ 2.\\
\begin{picture}(400,180)
\put(200,20){\makebox(0,0)[bc]{Figure 2: The integration contour in $Q_m$.}}
\end{picture}

For convenience, let
\beq
&z_1=z,&X(\{z_i\})=\prod_{i=1}^m(\psi^++\bt \psi)(z_i),\no\\
&k=[m/2],&i\df\phi_-(z)=\sum_{n<0}\phi_n/z^{n+1},
\eeq
where $[x]$ is the maximal integer no bigger than $x$.
The most singular term in the OPE of $X(\{z_i\})$ is
\beq
\frac{1}{2^kk!}\sum_{P\in S(m)}(-1)^{\pi(P)}\prod_{i=1}^k
\frac{\bt(z_{P_{2i-1}})+\bt(z_{P_{2i}})}{z_{P_{2i-1}}-z_{P_{2i}}}
(\psi^+(z_{P_m})+\beta\psi(z_{P_m}))^{m-2k},
\eeq
where $P$ is a permutation of the set $\{1,2,\ldots,m\},~\pi(P)$ its
$Z_2$ degree.
\beq
Q_m|m,-s\rangle=\oint \prod_{i=1}^m d\!z_i\prod_{1\leq i<j\leq m}
(z_i-z_j)^{\af_-^2}\prod_{i=1}^mz_i^{-2\af_-^2j_{m,-s}}
X(\{z_i\}) e^{i\af_-\sum_{i=1}^m\phi_-(z_i)}|-m,-s\rangle.
\label{9331101}
\eeq
Let $n=m(s-2)/2+k$, and
\beq
\langle\phi|&=&\langle -m,-s|\r^k_{1}\psi_{1}^{m-2k}\phi_{n}n/\af_-,
{}~~~~~\mbox{if }n\neq 0;\no\\
\langle\phi|&=&\langle -m,-s|\r^k_{1}\psi_{1}^{m-2k}m,~~~~~~~~~~~~~\mbox{if }
n= 0.
\eeq
Consider the inner product,
\beq
\langle\phi|Q_m|m,-s\rangle
&=&1/2^k\oint \prod_{i=1}^m d\!z_i\prod_{1\leq i<
j\leq m}(z_i-z_j)^{\af_-^2}\prod_{i=1}^m z_i^{-2\af_-^2j_{m,-s}}\no\\
&&\sum_{P\in S(m)}\prod_{i=1}^k(-1)^{\pi(P)}(z_{P_{2i-1}}-
z_{P_{2i}})^{-1}\sum_{i=1}^mz_i^n
\label{9331102}
\eeq
To see that the integration on the
r.h.s. of the eq.(\ref{9331102}) does not vanish in general,
it is more convenient to recast the integrand in a more suitable form.

\blm
Let $m>1$,
\beq
&&A_m(z_1,z_2,\cdots,z_m)=\prod_{1\leq i<j\leq m}(z_i-z_j),\no\\
&&f_m(z_1,z_2,\cdots,z_m)=\frac{A(z_1,z_2,\cdots,z_m)}{2^kk!}
\sum_{P\in S(m)}(-1)^{\pi(P)}\prod_{i=1}^k(z_{P_{2i-1}}-z_{P_{2i}})^{-1},\\
&&g_m(z_1,z_2,\cdots,z_m)=\frac{A^2(z_1,z_2,\cdots,z_m)}{ 2^kk!(m-k)!}
\sum_{P\in S(m)}(-1)^{\pi(P)}\prod_{i=1}^k\prod_{j=k+1}^{m}
(z_{P_{i}}-z_{P_{j}})^{-2},\no
\eeq
then
\beq
f_m(z_1,z_2,\cdots,z_m)=g_m(z_1,z_2,\cdots,z_m),~~~~\forall~m>1
\label{9331103}
\eeq
\elm
Proof. If $m=2,~3$, eq.(\ref{9331103}) can be verified by direct computation.\\
Let $m>3$, now we prove the lemma by induction on $m$. Assume that
\beq
f_k((z_1,z_2,\cdots,z_k)=g_k(z_1,z_2,\cdots,z_k),~\forall 1<k<m.
\eeq
Let us consider a particular case for which $z_{m-1}=z_m$. Then
\beq
f_m(z_1,z_2,\cdots,z_{m-1},z_{m})|_{z_m=z_{m-1}}
=\prod_{i=1}^{m-2}
(z_i-z_{m-1})^2f_{m-2}(z_1,z_2,\cdots,z_{m-2}),\no\\
g_m(z_1,z_2,\cdots,z_{m-1},z_{m})|_{z_m=z_{m-1}}
=\prod_{i=1}^{m-2}
(z_i-z_{m-1})^2g_{m-2}(z_1,z_2,\cdots,z_{m-2}).
\label{z1z2}\eeq
By induction
\beq
f_{m-2}(z_1,z_2,\cdots,z_{m-2})=g_{m-2}(z_1,z_2,\cdots,z_{m-2}).
\label{fg}\eeq
{}From eq.(\ref{z1z2}) and (\ref{fg}) we see that
\beq
f_m(z_1,z_2,\cdots,z_m)-g_m(z_1,z_2,\cdots,z_m)=0,~~~~\mbox{ if }z_{m-1}=z_m.
\eeq
So
\beq
(z_{m-1}-z_m)|(f_m(z_1,z_2,\cdots,z_m)-g_m(z_1,z_2,\cdots,z_m)).
\eeq
Secondly, from the fact that $f_m(\{z_i\})-g_m(\{z_i\})$ is a symmetric
homogeneous polynomial  of $(z_1,z_2,\ldots,z_m)$, we have
\beq
A_m(z_1,z_2,\cdots,z_m)|(f_m(z_1,z_2,\cdots,z_{m})-g_m(z_1,z_2,\cdots,z_{m}))
.\eeq
However, the degree of $A_m(z_1,z_2,\cdots,z_m)$ is $m(m-1)/2$, larger than
that of $f_m(\{z_i\})-g_m(\{z_i\})$, if $f_m\neq g_m$. So the only possibility
is that
\beq
f_m(z_1,z_2,\cdots,z_{m})=g_m(z_1,z_2,\cdots,z_{m}).
\eeq
This completes the proof of the lemma by the induction rule.

By lemma 4, eq.(\ref{9331102}) can be rewritten as
\beq
\langle\phi|Q_m|m,-s\rangle&=&
\oint_{C_i}\prod_{i=1}^mdz_i\prod_{1\leq i<j\leq m}
(z_i-z_j)^{\af_-^2-1}\prod_{i=1}^mz_i^{-2\af_-^2j_{m,-s}}g_m(z_1,z_2,\cdots,
z_m)\sum_{i=1}^mz_i^n\no\\
&=&2\pi i
\oint_{C_i}\prod_{i=2}^mdu_i\,S(u_2,\cdots,
u_m)\times \no\\
&&\times\prod_{2\leq i<j\leq m}
(u_i-u_j)^{\af_-^2-1}\prod_{i=2}^mu_i^{-2\af_-^2j_{m,-s}}
(1-u_i)^{\af_-^2-1},\label{ui}
\eeq
where $z_i=z_1u_i,~i=2,\ldots,m$, the integration over variable
$z_1$ is completed and
\beq
S(u_2,\cdots,u_m)=k!g_m(z_1,z_2,\cdots,z_m)\sum_{i=1}^mz_i^n\,
z_1^{-m(m+s-3)/2},
\eeq
is a symmetric polynomial function of $(u_2,\ldots,u_m)$.
To evaluate the integral on the r.h.s. of eq.(\ref{ui}),
let us first consider a more general case,
\beq
\oint _{C_i}\prod_{i=1}^n u_i^a (1-u_i)^b\prod_{1\leq i<j\leq n}(u_i-u_j)^c
S(u_1,u_2,\cdots,u_n),
\eeq
with the integral contour depicted as in fig.\ 3,
where $S(\{u_i\})$ is a symmetric polynomial of $\{u_i\}$.\\
\begin{picture}(400,180)
\put(200,20){\makebox(0,0)[bc]{Figure 3: The integration contour for
$\prod_i du_i$.}}
\end{picture}

Now we deform the integral contour to get an integration over $(1,-\infty)$.
For convenience, we introduce a notation.
\beq
J_r&=&\int_{c'_1}\cdots\int_{c'_{r-1}}\oint_{c_r}\oint_{c_n}\prod_{j=1}^nd_{u_j}
\prod_{j=1}^{r-1}(u_j-1)^b\prod_{j=r}^n(1-u_j)^b\no\\
&&\prod_{j=1}^nu_j^a\prod_{1\leq i<j\leq n}(u_i-u_j)^cS(u_1,u_2,\cdots,u_n),
\label{9331401}
\eeq
where the contours $C'_i,\ C_j$ are depicted in fig.\ 4.
When $a,\ b, c$ take general
values in the complex plane, $J_r$ should be considered as its analytic
continuation.\\
\begin{picture}(500,180)
\put(200,20){\makebox(0,0)[bc]{Figure 4: The deformed contour.}}
\end{picture}
As in the same approach to the calculation of Dotsenko-Fateev integration
\cite{DF}, we get a inductive relation between $J_r$'s.
\beq
J_r=e^{-i\pi b}(1-e^{i2\pi[(n-r)c+b+a]+i\pi(r-1)c})J_{r+1},\label{169}
\eeq
It is easy to get that
\beq
J_{n+1}&=&\int_{C'_i}\prod_{i=1}^ndu_{i}
\prod_{j=1}^n(u_j-1)^b
\prod_{j=1}^nu_j^a\prod_{1\leq i<j\leq n}(u_i-u_j)^cS(u_1,u_2,\cdots,u_n)
\no \\
&=&\frac{1}{n!}\prod_{j=1}^n(1+e^{-i\pi c}+\cdots+e^{-i\pi c(j-1)})
{\cal J},
\label{9331402}
\eeq
where
\beq
{\cal J}=\int_{1}^{\infty}\cdots\int_1^{\infty}\prod_{j=1}^ndu_j\prod_{j=1}^n
u_j^a(u_j-1)^b
\prod_{1\leq i<j\leq n}|u_i-u_j|^cS(u_1,u_2,\cdots,u_n).\label{9331403}
\eeq
Combinig eq.(\ref{169}) and  eq.(\ref{9331402}) we have
\beq
J_1&=&\frac{(-i)^n}{n!}e^{i\pi cn(n-1)/2+i\pi na}\left(\prod_{j=1}^n
\frac{\sin \pi((n-j/2-1)c+b+a)\sin (\pi cj/2)}{\sin(\pi c/2)}\right){\cal J},
\eeq
In our case
\beq
 a=-\frac{(m-1)\af_-^2+s}{2},~~~~~b=c=\af_-^2-1,
\eeq
The integrand in $\cal J$ is always positive definite
except for a measure zero set. It is easy to see that ${\cal J}\neq 0$. So
\beq
\langle\phi|Q_m|m,-s\rangle=\frac{2\pi i(-2i)^{m-1}}{(m-1)!}e^{-i\pi(m-1)c/2}
\left(\prod_{j=1}^{m-1}\frac{\sin^2\pi jc/2}{\sin\pi c/2}\right){\cal J}
\eeq
is not zero provided $0<m<\td{q}$.
Then the state $Q_m|m,-s\rangle$ is nonvanishing in eq.(\ref{9331102}).

Similarly another part of the proposition can be proved in the same way.

{\bf Figure Caption}\\
\vspace{1ex}\\
{}~~Fig.1: Boundary conditions on the torus.\\
{}~~Fig.2: The integration contour in $Q_m$.\\
{}~~Fig.3: The integration contour for $\prod_i du_i$.\\
{}~~Fig.4: The deformed contour.
\end{document}